\documentclass{aa}  
\usepackage{graphicx}
\usepackage{txfonts}
\usepackage{longtable}

\begin{document}
\title{The building up of the disk galaxy  M33 \\  and the evolution of the metallicity gradient}

\titlerunning{The evolution of M33}

\author{Laura Magrini, Edvige Corbelli, Daniele Galli}
\authorrunning{Magrini et al.}

\offprints{L. Magrini}

\institute{INAF--Osservatorio Astrofisico di Arcetri, Largo E. Fermi, 5, I-50125 Firenze, Italy\\
\email{laura@arcetri.astro.it}}

\date{Received ; accepted}

\abstract
{  The evolution of radial gradients of metallicity in disk galaxies
and its relation with the disk formation are not well understood.
Theoretical models of galactic chemical evolution make contrasting
predictions about the time evolution of metallicity gradients.}
{To test chemical evolution models and trace the star formation and
accretion history of low luminosity disk galaxies we focus on the Local
Group galaxy M33.}
{We analyze O/H and S/H abundances in planetary nebulae, H{\sc ii}
regions, and young stars, together with known [Fe/H] abundances in the
old stellar population of M33.  With a theoretical model, we follow
the time evolution of gas (diffuse and condensed in clouds), stars,
and chemical abundances in the disk of M33, assuming that the galaxy is
accreting gas from an external reservoir.}
{Our model is able to reproduce the available observational constraints
on the distribution of gas and stars in M33 and to predict the time
evolution of several chemical abundances.  In particular, we find that
a model characterized by a continuous infall of gas on the disk, at a
rate of $\dot M_{\rm inf}\approx 1$~$M_\odot$~yr$^{-1}$, almost
constant with time, can also account for the relatively high rate of
star formation and for the shallow chemical gradients.}
{Supported by a large sample of high resolution observations for this
nearby galaxy, we conclude that the metallicity in the disk of M33 has
increased with time at all radii, with a continuous flattening of the
gradient over the last $\sim 8$~Gyr.}


\maketitle
 
\section{Introduction}

Local Group (LG) galaxies, with their variety of morphological types,
are ideal candidates for testing the predictions of galactic chemical evolution
(GCE) models. The advent of a new generation of telescopes and sensitive
instruments, such as wide field cameras and high resolution spectrographs,
has given detailed information about the distribution of atomic and
molecular gas, stellar content, chemical abundances from H{\sc ii}
regions, planetary nebulae (PNe), and stars in external galaxies,
especially in the LG.

M33 (NGC~598) is a low-luminosity late-type spiral galaxy (morphological
type Sc II-III, cf. van den Bergh \cite{vdb00}) located in the LG at a
distance of 840~kpc (Freedman et al.~\cite{freedman91}).  Its proximity
implies a large angular size which, together with its modest inclination
($i=54^\circ$), makes this galaxy particularly suitable for detailed
studies of its stellar content, gas distribution, and abundance gradients.

M33 is an ideal target to test GCE models and  constrain
galaxy assembly processes because it shows no signs of recent
mergers and no presence of a prominent bulge or bar component
(Regan \& Vogel~\cite{regan94}).  If mergers were a recent dominant
process then an extended stellar halo with several substructures 
should be visible (Bullock \&
Johnston~\cite{bullock05}).  Spectroscopic surveys of red giant branch
(RGB) stars in M33 show no clear evidence for these substructures
(Ferguson et al.~\cite{ferguson06}), indicating that M33 is evolving
relatively undisturbed.  In addition, star counts in the outer disk
and individual stellar velocity dispersion measures in the optical disk
indicate that any stellar halo component contributes for less than
a few percent to the total luminosity (Barker et al.~\cite{barker06},
McConnachie et al.~\cite{mcconnachie06}).

The recent identification of individual stars in M33 (e.g. Block et
al.~\cite{block06}, Barker et al.~\cite{barker06}) suggests the existence
of different stellar populations and therefore of separate episodes of
star formation, but a satisfactory understanding of the overall evolution
of M33 is still lacking.  Chemical abundances of stellar populations
of different ages, fundamental constraints for GCE models, are becoming
available for this galaxy (e.g. Magrini et al.~\cite{m04}, \cite{m07};
Urbaneja et al.~\cite{urbaneja05}; Crockett et al.~\cite{crockett06};
Barker et al.~\cite{barker06}).  Surveys of molecular hydrogen through
the CO~J=1-0 line (Engargiola et al.~\cite{engargiola03}, Heyer et
al.~\cite{heyer04}), together with limits on stellar masses from dynamical
mass modeling (Corbelli~\cite{corbelli03}) and photometric surveys,
now allow a quantitative comparison between observable quantities and
model predictions.

Early attempts to model the chemical evolution of M33 date back to
the seminal work of Diaz \& Tosi~(\cite{diaz84}), and were later developed by
Moll\'a et al.~(\cite{molla96}) through a formalism originally
developed by Ferrini et al.~(\cite{ferrini92}).  Moll\'a
et al.~(\cite{molla96}) and Moll\'a et al.~(\cite{molla97}) 
tackled the issue of the time evolution of
radial distributions of elemental abundances in disk galaxies of
different morphological type, including M33, introducing in their model 
a radial dependence of the infall rate and the star formation rate (hereafter SFR).
A similar approach was also taken by 
Goetz \& Koeppen~(\cite{goetz92}), and Koeppen~(\cite{koeppen94}) 
with analytical GCE models.
More recently, Moll\'a \& Diaz (\cite{molla05}) parameterized inputs
and observational data relative to disk galaxies in order to compute a
grid of multiphase chemical evolution models able to predict the
evolution of irregular and spiral galaxies.  Their predictions compare
well (in a statistical sense) with the available data for several
nearby galaxies, including M33.
The models for M33 quoted above need however to be reconsidered at the 
light of the new data that have recently become available, such as radial
gas profiles and the metallicity determinations in stellar populations
of different ages (e.g. Barker et al.~\cite{barker06}, Magrini et
al.~\cite{m04}). The goal of this paper is therefore to model in 
detail the evolution of M33 in order to reach a satisfactory agreement
with the full set of available observations. In particular, we wish
to investigate the controversial issue of the time evolution of radial
metallicity gradients.

The problem of the time variation of radial chemical gradients
is far from being settled, either theoretically or observationally,
even in the case of our own Galaxy (see e.g. Goetz \& Koeppen~\cite{goetz92}; 
Koeppen~\cite{koeppen94}; Tosi~\cite{tosi96};
Moll{\'a} et al.~\cite{molla97}; Henry \& Worthey~\cite{henry99};
Maciel~\cite{maciel00}; Maciel et al.~\cite{maciel03}; Stanghellini
et al.~\cite{stanghellini06}). Different models predict opposite
behaviours of the metallicity gradient, showing the sensitivity of this
observable to the adopted parameterization of the physical processes.
In this respect, GCE models can be divided in two groups: those
where the metallicity gradients steepen with time (Tosi \cite{tosi88};
Chiappini et al.~\cite{chiappini97}; Chiappini et al.~\cite{chiappini01})
and those where they flatten with time (Moll\'a et al.~\cite{molla97};
Portinari \& Chiosi~\cite{portinari99}; Hou et al.~\cite{hou00}). The
main difference between the two groups is how fast chemical enrichment
proceeds in inner and outer regions of the galactic disk.  In models of
the former group, the outer disk is pre-enriched by the halo, and its
metallicity is affected very little by the subsequent low star formation
activity. In contrast, in models of the latter group, a vigorous star
formation activity in the inner disk results in a rapid increase of the
metallicity near the galactic center.  Here the metallicity increases very
fast reaching its final value in the first 2--3 Gyr of disk evolution,
whereas the outer disk enrichment increases much slowly and progressively
flattens the radial gradient.

Observationally, it is possible to discriminate between these two groups
of models by comparing the metallicity gradient of the ``old'' 
stellar population (RGB stars and PNe) with the present-day gradient
outlined by H{\sc ii} regions and young stars.  In particular, the
comparison of PNe and H{\sc ii} regions  investigates the similarities
between the observational techniques and abundance analysis for
the two samples (Maciel et al.~\cite{maciel05}, \cite{maciel06}).
In our Galaxy the time variation of the gradients is mainly
derived from chemical abundances of PNe, open clusters, Cepheids and
young objects, such as OB associations of stars and H{\sc ii} regions
(e.g. Maciel et al.~\cite{maciel03}, \cite{maciel05}, \cite{maciel06};
Friel~\cite{friel95}; Friel et al.~\cite{friel02}). Observationally,
the [Fe/H] gradient in open clusters supports a time flattening  of the
abundance gradient
in agreement with the gradients derived from Cepheids of well
determined age.  According to Maciel et al.~(\cite{maciel05}), the
average flattening rate for the last 8 Gyr is about 0.005--0.010
dex~kpc$^{-1}$Gyr$^{-1}$.

In M33, it is not possible to derive the time evolution of the
gradient from abundances of PNe alone, as in the case of the our own Galaxy,
because of the difficulty to sub-divide PNe in different age groups
(Magrini et al.~\cite{m04}).  However, a comparison between abundances
measured in H{\sc ii} regions (tracing present-day abundances in the
interstellar medium, ISM) and in PNe (tracing the ISM composition at the
time of formation of their progenitors) should in principle reveal any
change in the radial distribution of the elements occurred in the last few
Gyr, for those elements not affected by low-mass stellar nucleosynthesis
like O, S and Ne.

The paper is organized as follows: in Sect.~\ref{sect_data} we collect
and critically examine the observational data adopted to constrain our
GCE model described in Sect.~\ref{sect_model}; our results are presented
in Sect.~\ref{sect_results}, with a particular attention to the issue of
the time evolution of the metallicity gradients (Sect.~\ref{sect_grad});
finally, in Sect.~\ref{sect_conc} we summarize our conclusions.

\section{Observational data}
\label{sect_data}

In this section we examine the observational constraints for the 
evolutionary model of M33, namely: the radial profiles of gas and stars,
the SFR, and the chemical abundances.

\subsection{Gas surface density}
\label{data_gas}

The gas and stellar mass surface densities in M33 have been  analyzed
by Corbelli (\cite{corbelli03}) using the gas kinematics.  The neutral
hydrogen surface density profile has been derived from high sensitivity
observations with the Arecibo-305m telescope after applying corrections
for the disk inclination with respect to the line-of-sight according
to the tilted ring model described in detail by Corbelli \& Salucci
(\cite{corbelli00}).

Imaging of molecular clouds complexes has recently been carried out
in M33 with the BIMA interferometer and with the FCRAO-14m telescope using
the CO~J=1-0 line transition (Corbelli~\cite{corbelli03}; Engargiola et
al.~\cite{engargiola03}; Heyer et al.~\cite{heyer04}). These data sets
reveal the presence of both compact, massive molecular clouds, and of
a more distributed molecular gas component. The total molecular mass
is $\sim 2\times 10^8$~$M_\odot$, corresponding to $\sim 10$\% of the
ISM mass of this galaxy.  Giant molecular clouds (GMCs) detected by
the interferometer account for only $\sim 20$\% of the molecular gas
mass. Most likely some of the total molecular mass detected by FCRAO
just resides around the complexes imaged by the interferometer while
some is in diffuse form spread out throughout a more extended area.

The azimuthally averaged surface density of atomic and molecular gas,
corrected for the disk inclination with respect to the line-of-sight, from data 
in Corbelli~(\cite{corbelli03}), can be well represented as follows:
\begin{equation}
\label{eq:gas}
\Sigma_{\rm gas} = 21 
\exp\left[-\left(\frac{R}{7.8~\mbox{kpc}}\right)^2
\right]~\mbox{$M_\odot$~pc$^{-2}$}.
\end{equation}

\subsection{Stellar surface density}
\label{data_star}

The main stellar component of M33 is a thin disk: the central bulge
and the stellar halo contain less than 10\% of the total luminous mass.
$K$-band photometry, which more closely reflects the underlying
stellar mass distribution in the disk being less affected by
extinction, points out to an exponential stellar light distribution
declining radially with a scale length of 1.4~kpc (Regan \&
Vogel~\cite{regan94}). Both $B$- and $K$-band photometry show that
departures of the stellar light profile from an exponential law are
visible in the innermost kpc, where either the disk
scale length changes, or a compact bulge is present. Using the
$K$-band scale length, together with data on the atomic and molecular
gas distribution, Corbelli~(\cite{corbelli03}) has inferred the
stellar mass-to-light ratio, and therefore the stellar mass density of
this galaxy. This includes also the mass in stellar remnants, being 
derived via a dynamical mass model. The best fitting three
component model (gas, stars, dark matter) to the rotation curve
requires a stellar surface mass density of
\begin{equation}
\label{eq:stars}
\Sigma_{\rm stars}=430\exp
\left[-\left(\frac{R}{1.42~\rm{kpc}}\right)\right]~\mbox{$M_\odot$~pc$^{-2}$}.
\end{equation}
The rotation curve of M33 rises in the center out to a galactocentric
radius $R \approx 3$~kpc, then it flattens out up to the outermost
measured point. Radial mixing flows, possibly driven by shear generated
by differential rotation, can be present at $R>3$~kpc.

\subsection{The SFR}
\label{sec_sfr}

The blue colour of M33 ($B-V=0.55$) and the prominent H{\sc ii} regions
indicate vigorous star formation activity in the nuclear region and
along the spiral arms, where the largest molecular complexes are found
(Engargiola et al.~\cite{engargiola03}).  In general, optical and
infrared luminosities can be used to trace the SFR in external galaxies.
Optical observations rely on the conversion of H$\alpha$ luminosities
into SFRs (Kennicutt~\cite{kennicutt98}).  The big uncertainty with this
method is the H$\alpha$ luminosity correction for dust extinction. Other
methods involving the far infrared (FIR) luminosity are less affected by
extinction.  In this case, however,  the conversion factor for deriving
the SFR depends sensitively on the assumption about the fraction of
the total luminosity re-irradiated in the infrared (IR) and also on the
type of stars contributing to dust heating. The end result is that the
IR conversion factor is subject to large uncertainties.  Fortunately,
M33 is known to have moderate extinction and the optical method seems
appropriate for this galaxy.  Extinction for M33 has been studied
by Israel \& Kennicutt~(\cite{israel80}) using a sample of bright
H{\sc ii} regions and free-free radio emission measurements. They
found  a clear radial dependence of the extinction law.  Devereux et
al.~(\cite{devereux97}) on a larger sample of H{\sc ii} region, using
the same radio-H$\alpha$ method, found no clear radial trend but an
average extinction of $A_v\approx 1$~mag.  We will use this value without
introducing any other ``uncertain'' radial dependence.

We determine the H$\alpha$ surface brightness profile from the
H$\alpha$ image of Hoopes \& Walterbos (\cite{hoopes00}), averaging the
H$\alpha$ flux along ellipses. These are projections of circular rings
0.5~kpc wide, inclined by $54^\circ$ at a position angle $23^\circ$.
The diffuse ionized gas (DIG) contributes for $\sim 40$\% to the total
observed H$\alpha$ flux  (Hoopes \& Walterbos~\cite{hoopes00}) and we
assume that the DIG emission is unaffected by dust extinction.  DIG
emission is however important in computing the SFR since field O and B
stars are responsible for it.  After subtracting 5\% of the recovered
flux due to possible [N{\sc ii}] lines contamination, and correcting
60\% of the emission for 0.83~mag of extinction (a factor 1.2 smaller
than extinction in the visual), the surface brightness is converted in
SFR density following Kennicutt et al.~(\cite{kennicutt98}),
\begin{equation}
{\rm SFR} = 
\frac{L({\rm H}\alpha)}{1.26\times 10^{41}~\mbox{erg~s$^{-1}$}}~\mbox{$M_\odot$~yr$^{-1}$}
\end{equation}

The resulting SFR per unit area  is shown in Fig.~\ref{Fig_sfr}.  The
SFR integrated over the whole disk is $\sim 0.4$~$M_\odot$~yr$^{-1}$,
comparable to the value given by Kennicutt (\cite{kennicutt98}) and
also to the value derived from the FIR luminosities (see Hippelein et
al.~\cite{hippelein03} and references therein). We shall use the SFR
derived by Kennicutt~(\cite{kennicutt89}) from H$\alpha$ luminosities
of bright H{\sc ii} complexes and the FIR SFR by Heyer at
al.(\cite{heyer04}) respectively as lower and upper limits for our
model results.

The SFR in the range from 1 to 8~Gyr ago  can be estimated from the PNe
population as described by Magrini et al.~(\cite{m05}). In short, the
method consists in counting PNe, estimating the mean mass of the PNe
progenitors and obtaining  a measure of the total mass of the
intermediate-age stellar population.  As a result, the intermediate-age
stellar mass is proportional to the number of PNe. Extrapolating the
number of known PNe (Ciardullo et al.~\cite{ciardullo04}) in M33 within
4 mag from the cut-off of their luminosity function (Ciardullo et
al.~\cite{ciardullo89}), and considering a time interval of 3~Gyr for
the formation of their progenitors (the youngest, since we are
considering the brightest PNe) we derive a mean
SFR=0.55~$M_\odot$~yr$^{-1}$,(1--4~Gyr ago) integrated over the whole
disk.  On the other hand, considering older PNe, extrapolating within 8
mag from the cut-off of their standard luminosity function, and
considering the whole time interval for the formation of progenitor
stars, we obtain a higher mean SFR=1.1~$M_\odot$~yr$^{-1}$ (1--8~Gyr
ago).

\begin{figure}
\centering
\resizebox{\hsize}{!}{\includegraphics[angle=0]{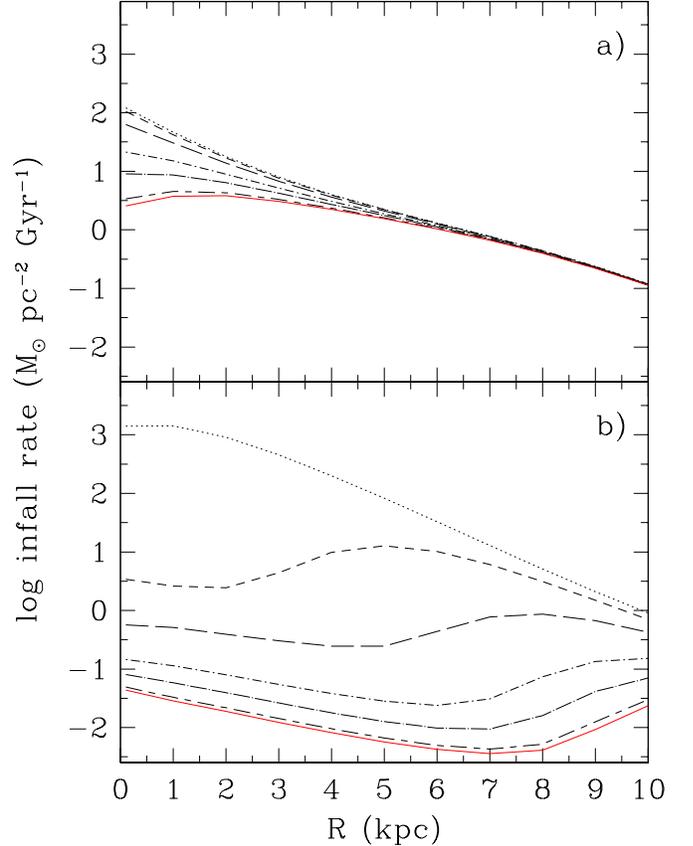}}
\caption{The time evolution of the infall rate  for the accretion model 
(panel a) and for the
collapse model (panel b).
Curves represents the infall rate at 0.5 ({\em dotted curve}), 
2 ({\em dashed curve}), 3 ({\em long-dashed curve}), 5 ({\em dot-dashed curve}), 
8 ({\em long dash-dotted curve}), 12 ({\em long-short dashed curve}), 
and at 13.6~Gyr ({\em solid curve}). }
\label{Fig_infall}
\end{figure}

\begin{figure}
\centering
\resizebox{\hsize}{!}{\includegraphics[angle=0]{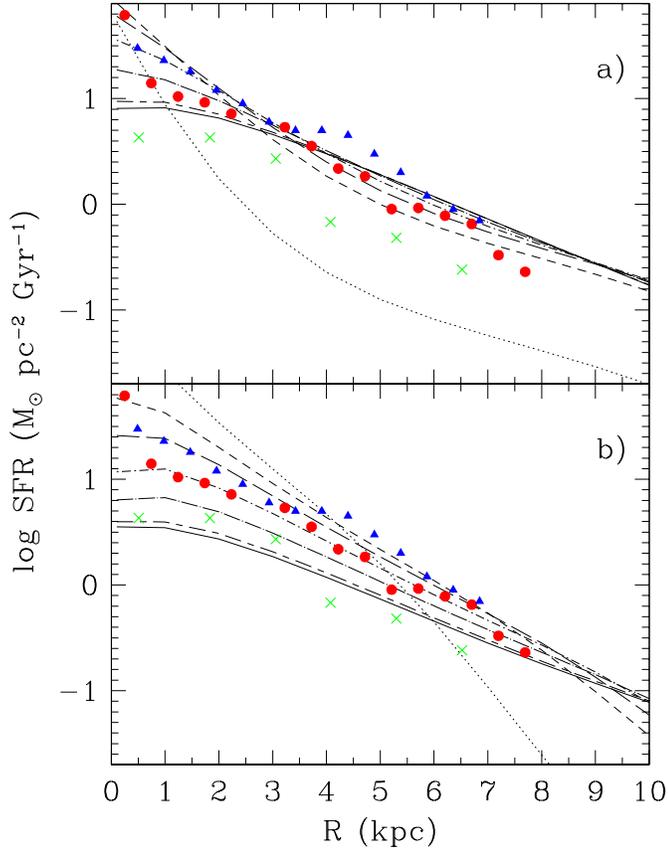}}
\caption{Comparison of  the predicted time evolution star formation rate 
(line types as in Fig.\ref{Fig_infall}) with
observations for the accretion model
(panel a) and with the collapse model (panel b).
Data are from  Kennicutt et al.~(\cite{kennicutt89}, {\em crosses}); 
Hoopes \& Walterbos~(\cite{hoopes00}, {\em filled circles}); 
Heyer et al.~(\cite{heyer04}, {\em triangles}).}
\label{Fig_sfr}
\end{figure}

\subsection{The infall rate}
\label{sec_obs_infall}

Star formation in M33 extends much beyond the region where molecular
complexes are located. Carbon stars have been found even at larger
radii where the drop of the H$\alpha$ flux occurs ($R\approx 6$~kpc,
Kennicutt \cite{kennicutt89}).  The presence of such an
intermediate-age stellar population  in the outer disk suggests that
accretion of gas is still taking place (Davidge~\cite{davidge03}; Block
et al.~\cite{block04}, \cite{block06}; Rowe et al.~\cite{rowe05}) and
that the disk of the galaxy is forming inside-out. A likely source of
the accreted gas could be the cosmic web (Keres et al.~\cite{keres05}),
the warm/hot filaments extending between galaxies, where most of the
baryons reside.

Is there any evidence that M33 still in the process of accreting mass
from the intergalactic medium? A conspicuous cloud of neutral gas
infalling into the disk of M33 recently detected at 21-cm by Westmeier
et al.~(\cite{westmeier05}) supports this picture. The cloud might have
been tidally stripped from a distant faint dwarf or dark galaxy (with
no associated stellar counterpart) or it might just be gas condensing
out of intergalactic medium filaments. This finding confirms that M33
is still in the process of accreting gas to fuel star formation. 
The gas infall rate over a disk region corresponding to the cloud
size is about 2~$M_\odot$~pc$^{-2}$~Gyr$^{-1}$. 
There are however observational difficulties
in estimating the total amount of gas infalling into the M33 disk. 
Velocities of the infalling clouds are expected to be similar to
velocities of high velocity clouds related to the MW disk, and 
therefore confusion limits our ability to distinguish the two
different populations in the small sky area where M33 is located.
Any high velocity cloud in projected proximity 
to the M33 disk, placed at the distance of M33 would result in a
much higher infall rate over M33 than what could be inferred 
for the same cloud if it were closer to us and infalling into 
the Milky Way (MW) disk. However, the number of high velocity clouds 
throughout the whole sky is large, and most of them are 
connected with gas infalling into the MW galaxy. Observational estimates 
of the infall rates into the MW disk range between 0.1 and 
5~$M_\odot$~yr$^{-1}$ (e.g.  Wakker et al.~\cite{wakker99}; 
de Boer~\cite{deboer04}). Rates  $\approx$~1~$M_\odot$~yr$^{-1}$
or larger are favoured by considerations on the rapid molecular gas depletion, 
on the G-dwarf problem, and by some recent evolution models of disc 
galaxies (e.g. Naab \& Ostriker \cite{naab06} and references therein).  
The total infall rate for M33 is expected to be lower than in the MW  
because its total mass is lower than the mass of the MW. This 
consideration applies both to halo gas infall models, which involve the
collapse time scale, and to models where the gas starts flowing in from the 
intergalactic medium (see Section 3.1).
Given that quantitative observations of the infall rate over a wide area of M33 and
a wide interval of time are difficult to obtain, in Sect.~\ref{sect_results} 
we shall discuss the implications of models characterized by different infall
rates.

\subsection{Chemical abundances: stars and ISM}
\label{sect_data_chem}

Aller~(\cite{aller42}) first obtained spectra of H{\sc ii} regions in
M33 and derived a radial gradient of the  [O{\sc ii}]/[O{\sc iii}] line
ratio, that he attributed to an excitation gradient.
Smith~(\cite{smith75}), in subsequent spectroscopic study, found clear
evidence for a metallicity gradient.  Further spectroscopic studies of
H{\sc ii} regions were carried on by Kwitter \&
Aller~(\cite{kwitter81}) and Vilchez et al.~(\cite{vilchez88}).
Garnett et al.~(\cite{garnett97}) using data from previous observations
with derived electron temperatures, re-determined the abundances
obtaining an overall O/H gradient of $-0.11 \pm 0.02$~dex~kpc$^{-1}$,
including the central regions of M33. The steepness of the
gradient is however dominated by the innermost data point. Subsequent
studies of the metallicity gradient which do not include the central
regions obtain a much shallower gradient. Recently, Willner \&
Nelson-Patel~(\cite{willner02}) have derived Ne/H abundances for 25
H{\sc ii} regions from infrared lines, obtaining a [Ne/H] gradient
$-0.034 \pm 0.015$~dex~kpc$^{-1}$, outside $\sim$0.5~kpc from the
center.  Crockett et al.~(\cite{crockett06}), measuring chemical
abundances in six H{\sc ii} regions, have derived a [Ne/H] gradient of
$-0.016 \pm 0.017$~dex~kpc$^{-1}$ and a [O/H] gradient of $-0.012 \pm
0.011$~dex~kpc$^{-1}$, much shallower than in previous studies, 
but at the limit of significance due to the smallness of the sample.
All these determinations are consistent with each other and imply
that excluding the innermost $\sim$0.5~kpc region where a bulge
might be present, the metallicity gradient  of H{\sc ii} regions in the
disk of M33 is very shallow.

Chemical abundances of PNe have been derived  by Magrini et
al.~(\cite{m04}) via optical spectroscopy and photoionization
modeling.  Stellar abundances have been obtained by Herrero et
al.~(\cite{herrero94}) for AB-type supergiant stars, McCarthy et
al.~(\cite{mccarthy95}) and Venn et al.~(\cite{venn98}) for A-type
supergiant stars, Monteverde et al. (\cite{monteverde97},
\cite{monteverde00}) and Urbaneja et al.~(\cite{urbaneja05}) for B-type
supergiant stars. The larger sample by Urbaneja et
al.~(\cite{urbaneja05}) indicates an [O/H] gradient of $-0.06 \pm
0.02$~dex~kpc$^{-1}$.  Very recently the detection of beat Cepheids,
which are young intermediate mass stars (Beaulieu et
al.~\cite{beaulieu06}) allowed the derivation of their metallicity
making use of stellar pulsation models and of the mass-luminosity
relation. Beaulieu et al.~(\cite{beaulieu06}) derived a metallicity
gradient of $-0.20$~dex~kpc$^{-1}$, and inferred an [O/H] gradient
$-0.16$~dex~kpc$^{-1}$.  Metallicities of the old stellar population
have been obtained via deep CCD photometry and colour-magnitude
diagrams by Stephens \& Frogel (\cite{stephens02}), Kim et al.
(\cite{kim02}) in inner fields, and by Galletti et al.
(\cite{galletti04}), Tiede et al.~(\cite{tiede04}), Brooks et al.
(\cite{brooks04}) and Barker et al.~(\cite{barker06}) in outer fields.
The [Fe/H] gradient related to the whole RGB stellar population has a slope
$-0.07 \pm 0.01$~dex~kpc$^{-1}$ (Barker et al.~\cite{barker06}).

\section{The model}
\label{sect_model}

The GCE model adopted in this work is a generalization of the
multi-phase model by Ferrini et al.~(\cite{ferrini92}), built for the
solar neighborhood, and subsequently extended to the entire MW (Ferrini
et al.~\cite{ferrini94}), and to other disk galaxies (e.g. Moll\'a et
al.~\cite{molla96}, \cite{molla97}; Moll\'a \& Diaz~\cite{molla05}).
Here we summarize the main characteristics of the model (see Ferrini et
al.~\cite{ferrini92} for details).

The galaxy is divided into $N$ coaxial cylindrical annuli with inner
and outer galactocentric radii $R_i$ ($i=1,N$) and $R_{i+1}$,
respectively, mean radius $R_{i+1/2}\equiv (R_i+R_{i+1})/2$ and height
$h(R_{i+1/2})$.  Each annulus is divided into two {\em zones}, the {\em
halo} and the {\em disk}, made of  diffuse gas $g$, clouds $c$, stars
$s$ and stellar remnants $r$.  The {\em halo} component is
 intended here quite generally as the primordial baryonic halo, or as
material accreted from interactions with small LG galaxies or from the
intergalactic medium during the life-time of the galaxy.  In the
following, a subscript $H$ or $D$ indicates the halo or the disk,
respectively.  For a spherical halo of radius $R_{N+1}\equiv R_H$,
\begin{equation}
h(R_{i+1/2})=(R_H^2-R_{i+1/2}^2)^{1/2},
\end{equation}
and the halo volume in the $i$--th annulus is then  
\begin{equation}
V_{H,i}=\pi (R_{i+1}^2-R_i^2) h(R_{i+1/2}).
\end{equation}
If $z_D$ is the disk scale height (assumed independent on
radius), the volume of the disk in each annulus is
\begin{equation}
V_{D,i}=\pi (R_{i+1}^2-R_i^2) z_D.
\end{equation}
At time $t=0$ all the mass of the galaxy is in the form of diffuse gas
in the halo. At later times, the mass fraction in the various components is
modified by several conversion processes: diffuse gas is converted into
clouds, clouds collapse to form stars and are disrupted by massive
stars, stars evolve into remnants and return a fraction of their mass
to the diffuse gas. The disk of mass $M_D(t)$ is formed by continuous
infall from the halo of mass $M_H(t)$ at a rate
\begin{equation}
\frac{{\rm d}M_D}{{\rm d}t}=fM_H,
\end{equation}
where $f$ is a coefficient of the order of the inverse of the infall
time scale. Clouds condense out of diffuse gas at a rate $\mu$ and 
are disrupted by cloud-cloud
collisions at a rate $H^\prime$,
\begin{equation}
\frac{{\rm d}M_c}{{\rm d}t}=\mu M_g^{3/2}-H^\prime M_c^2,
\end{equation}
where $M_g(t)$ and $M_c(t)$ are the mass fractions of diffuse gas and
clouds, respectively.  Stars form in the halo and the disk by
cloud-cloud collisions at a rate $H$ and by the interactions of massive
stars with clouds at a rate $a$,
\begin{equation}
\frac{{\rm d}M_s}{{\rm d}t}=H M_c^2+aM_sM_c-DM_s,
\end{equation}
where $M_s(t)$ is the mass fraction in stars and $D$ is the stellar death rate.

Each annulus is then evolved independently (i.e. without radial mass flows)
keeping fixed its total mass from $t=0$ to $t_{\rm gal}=13.6$~Gyr
computing the fraction of mass in each component in the two zones, and
the chemical composition of the gas (assumed identical for diffuse gas
and clouds).  The rate coefficients of the model are all assumed to be
independent on time but functions of the galactocentric radius $R$.
Their radial dependence for a model of the MW is discussed in
detail by Ferrini et al.~(\cite{ferrini94}), Moll\'a et
al.~(\cite{molla96}), and Moll\'a \& Diaz~(\cite{molla05}).  In
general, coefficients representing condensation processes (like e.g.
the formation of clouds from diffuse gas), being proportional to the
inverse of the dynamical time, scale with the inverse square root of
the zone volume, whereas the coefficients of binary processes (like
e.g. the formation of stars by cloud-cloud collisions) scale as the
inverse of the zone volume; the coefficient of star formation induced
by stars is independent on radius.

\begin{table}
\label{table:m33}      
\begin{tabular}{ll}        
\hline
Distance              & 840~kpc            \\
Type                  & Sc II-III                \\
Inclination           & $54^\circ$         \\
Position angle        & $23^\circ$         \\
Luminosity            & $6.5\times 10^9$~$L_\odot$ \\
Maximum rotation speed       & 130~km~s$^{-1}$    \\
Optical radius $R_{\rm opt}$ & 6.6~kpc            \\
Equivalent Solar radius      & 3.9~kpc            \\
\hline
\end{tabular}
\caption{M33: adopted physical characteristics}            
\end{table}

The model described in this section can be applied in general to study
the chemical evolution of any disk galaxy. We now discuss the values of
the various coefficients that we have specifically adopted for M33.  As
a starting point, we have adopted  the set of values by Ferrini et
al.~(\cite{ferrini92}) for their ``best model'' of the solar
neighborhood, applying them at the equivalent solar radius
\footnote{defined as $R_{\odot,{\rm M33}}=(R_{\rm opt,M33}/R_{\rm
{opt,MW}})R_{\odot,{\rm MW}}$}. The values at any other radius are then
determined by the scaling relations described in Sect.~\ref{sect_model}.

The mass fractions in each component computed by the model in each
annulus are then converted into surface densities using as a
normalization the total  surface density, (i.e. the sum of the H{\sc
i}, H$_2$, stellar and remnants surface densities,  as given by
equations~\ref{eq:gas} and \ref{eq:stars}).  In the comparison with
the data, we identify the gas and  cloud components  with the H{\sc i}
and the H$_2$ gas, respectively.

\subsection{The infall coefficient} 
\label{sec_infall}

We assume an exponential  radial dependence of the infall rate $f$,
\begin{equation}
f(R) = f_0 \exp(-R/\lambda_D)
\end{equation} 
where $f_0$ is the infall rate at $R=0$ and $\lambda_D$ is a typical
disk scale length (cf. Ferrini et al. \cite{ferrini94}, Portinari \&
Chiosi \cite{portinari99}).  In principle, $\lambda_D$ can be set equal
to the scale length of the disk surface brightness in the $B$-band
($\sim 1.9$~kpc, Freeman \cite{freeman70}), or in the $K$-band ($\sim
1.4$~kpc, Regan \& Vogel \cite{regan94}), or to the CO scale length
($\sim 2.7$~kpc, Corbelli \cite{corbelli03}).  In our model, the value
of $\lambda_D$ that better reproduces the observations (gas and stars
surface density profiles, radial SFR, see Figs.  \ref{Fig_sfr},
\ref{Fig_stars}, \ref{Fig_hi}, \ref{Fig_hii}) is $\lambda_D \approx
2.0$~kpc.  Larger values of $\lambda_D$ imply a huge amount of neutral
hydrogen in the central regions, while lower values depress the SFR in
the inner regions.

The infall coefficient $f_0$ is usually set of the order of the inverse
of the collapse time $\tau_c$ for the galaxy, 
where $\tau_c$ depends on the total mass of the galaxy $M_{\rm Gal}$
as $\tau_c\propto M_{\rm Gal}^{-1/2}$ (Gallagher et
al.~\cite{gallagher84}).  This choice corresponds to a scenario where
the disk is formed very rapidly at the beginning of a galaxy's life
and we shall refer to this model as the {\em collapse} model.  Since
the mass of M33 is about 0.2 times the mass of the MW, the value of
$f_0$ for M33 should be a factor $\sim 2.2$ smaller than in the case
of the MW.  Scaling the infall value adopted by Ferrini et
al.~(\cite{ferrini92}) for the MW ($f_0=0.7$ see Table~\ref{Tab_model}),
corresponding to a dynamical collapse time of the order of 10$^8$~yr
at the solar radius, we obtain for M33 $f_0\approx 0.3$.  Over a disk
of 10~kpc in radius this choice results in a present-day average
accretion rate of $\dot M_{\rm inf}\approx 0.05
$~$M_\odot$~pc$^{-2}$~Gyr$^{-1}$ (Fig~\ref{Fig_infall}b). 
Thus our {\em collapse model} is intended to reproduce the 
characteristics of a rapid phase of disk formation, following the 
dynamical collapse of an extended halo. 

However, estimates of $f_0$ for the
MW based on chemical constraints (G-dwarfs metallicity distribution,
O/Fe vs. Fe/H relations, etc.)  suggest lower values, $f_0\approx
0.012$ for the MW, corresponding to a time scale of the order of several 
Gyr (see Moll\'a \& Diaz~\cite{molla05} and references therein). Lower values
of $f_0$ implies higher infall rates at the present time. If this is
the case for M33 it is unlikely however that the halo is the reservoir of  
the gas infalling into the disk because there are no observational evidences
for a gaseous halo.
A possibility for supporting a high infall rate at present time, and
in general an infall rate almost constant in time, is to consider a  
slow continuous accretion process from the environment.  This
is obtained for example using $f_0\approx 0.003$, corresponding to an
infall rate of 1.2~M$_\odot$~yr$^{-1}$ in the M33 disk at present time or
about and average of $\dot M_{\rm inf}\approx
3.8$~$M_\odot$~pc$^{-2}$~Gyr$^{-1}$ (Fig~\ref{Fig_infall}a).
We shall refer to this model as the {\em accretion} model.

High infall rates are favoured by some recent papers
describing the evolution of disk galaxies (e.g. Naab \&
Ostriker~\cite{naab06}). Infall rate in the MW for example are
predicted to be of order 2--4~M$_\odot$~yr$^{-1}$, with little variations
over the last 10~Gyr.  These values are in good agreement with those
inferred from observations of high velocity clouds and explain 
specific abundance patterns such as the high deuterium abundance at the
solar neighborhood and at the Galactic Center (Chiappini et
al.~\cite{chiappini02}). In these models the intergalactic medium is
often identified as the source of the gas infalling into the disk.
Hydrodynamical simulations of galaxy formation and evolution
predict that accretion of `cold' gas from intergalactic filaments is 
present and becomes the dominant process for low mass galaxies and in 
low density regions. Galaxies of mass similar to M33 should accrete gas 
at a rate which decreases slowly with time and is currently
$\approx$~1~$M_\odot$~yr$^{-1}$ (Keres et al.~\cite{keres05}). 
This finding supports the infall rates used in our {\em accretion} model. 
Higher mass halos have larger accretion
rates but a large fraction of the accreted gas has been shock heated at
the time the galaxy formed, and does not flow in from large distances,
as in the case of intergalactic filaments feeding the disk.  The
intergalactic feeding mechanism is well suited for M33 because of the
absence of a massive gaseous and stellar halo, relic of the galaxy 
formation epoch and driver of disk accretion processes.

\subsection{The star and cloud formation efficiencies}
\label{sec_eff}

The parameters regulating the conversion of diffuse gas into clouds
and clouds into stars are: $\mu$ (cloud formation from diffuse gas),
$H$ (star formation from cloud collisions), and $H^\prime$ (cloud
dispersion).  The efficiencies of these processes adopted in our
models, together with previous values adopted for the MW and M33, are
shown in Table~\ref{Tab_model}. These efficiencies are in general
functions of the morphological type of the galaxy (cf. Sandage
\cite{sandage86}; Gallagher et al.~\cite{gallagher84}); Ferrini \&
Galli~(\cite{ferrini88}) analyzed the behaviour of these parameters in
the different morphological types of spiral galaxies, and found a
reduction of $\mu$ and $H$ from a Sbc galaxy, like the MW, to a Scd
galaxy, like M33.  These scaling relations can be also compared with
those of Moll\'a \& Diaz (\cite{molla05}) who studied spiral galaxies
of different morphological types.  For a spiral galaxy of
morphological type 6, such as M33, they adopted the following values
(see their Table~2): $\epsilon_H=0.01$ and $\epsilon_\mu=0.15$, which
translate,  at the equivalent solar radius, into  $H=0.012$
and $\mu =0.017$ in units of 10$^{-7}$~yr$^{-1}$, using the relations between
$\epsilon_H$, and $H$ and between $\epsilon_\mu$ and $\mu$ given by
Ferrini et al.~(\cite{ferrini94}).
The  $H$ parameter adopted in the present work, in particular for
the {\em accretion} model, has a higher value than that used for M33
by Moll\'a \& Diaz~(\cite{molla05}) and by Moll\`a et
al.~(\cite{molla96},\cite{molla97}).  This is 
probably due to the different parameterization of the IMF used:
Moll\`a and collaborators used the Ferrini's et
al.~(\cite{ferrini90}) IMF which predicts a larger number of massive
stars with respect to other IMF, such as the Kroupa's et al.~(\cite{kroupa93})
IMF.
Therefore their model needed a lower $H$ value in order to reproduce the observed 
metal content.

Note that in order to reproduce the observed lower content of molecular gas
(Fig.~\ref{Fig_hii}) of M33 with respect to that found in Sc galaxies,
a very low cloud formation efficiency is needed.  The low molecular fraction
in this galaxy is in agreement with the tendency
of the gas fraction in molecular form to decrease from Sc to Irr
morphological types, despite the higher SFR of the latter.

\subsection{The cloud dispersal coefficient}

The cloud dispersal coefficient $H^\prime$ is a measure of the
probability that collisions between gas clouds result in the disruption
of the clouds themselves and return cloud material to the
diffuse phase. In principle, the rate of cloud dispersal should depend
on radius in the same way as the rate of star formation by cloud-cloud
collisions, namely as the inverse of the disk volume $V_D$ (see Ferrini
et al.~\cite{ferrini94}).  However, this dependence makes impossible to
reproduce the observed distributions of H{\sc i} and H$_2$, because at
small galactocentric radii, where the factor $1/V_D$ enhances the rate
of formation of diffuse gas by cloud collisions, the H{\sc i} gas is
underabundant (see Fig.~\ref{Fig_hi}). Our model requires instead that
the coefficient of cloud dispersal, $H^\prime$, should be roughly
independent on radius.  Since the dependence of the frequency of
collisions from the inverse of the volume is a general characteristic
of any binary collision process, an additional hypothesis must be made
about the efficiency of the process of cloud dispersal.  Namely, our
modeling requires that the clouds in the outer galaxy are more
efficiently dispersed by collisions than clouds in the inner galaxy.

This model requirement is related to the observed absence of GMCs
at large radii where star formation is still taking place, supporting
the claim that clouds are easily destroyed in the outer disk of M33.
This may be due to a radial change of some intrinsic property of the
clouds (e.g. compactness) or to additional processes that regulate the
molecular/atomic hydrogen ratio in galaxies of low molecular fraction
such as M33, like photodissociation by a pervasive interstellar
radiation field (Elmegreen \cite{elmegreen93}; Heyer et
al.~\cite{heyer04}). Because of the larger photon mean free path at
larger radii, this process probably dominates the conversion of
molecular clouds into atomic diffuse gas in the outer disk of M33.

\subsection{Stellar  yields}
\label{sec_cheyields}

We model the chemical enrichment of the gas using the formalism
developed by Talbot \& Arnett~(\cite{talbot73}) who introduced the
restitution matrices $Q_{i,j}(M,Z)$.  The elements of these matrices
are defined as the fraction of the mass of an element $j$ initially
present in a star of mass $M$ and metallicity $Z$ that it is converted
in an element $i$ and ejected. At each stellar mass $M$ and metallicity
$Z$ corresponds one matrix $Q_{i,j}(M,Z)$.  Our GCE model takes into
account two different metallicities, $Z=0.02$ and $Z=0.006$, and 22 
stellar masses (21 for $Z=0.006$), ranging from 0.8 to 100~$M_\odot$, for
a total of 43 restitution matrices.

We update the nucleosynthesis yields used to build the $Q_{i,j}(M,Z)$
matrices in the following way. For low- and intermediate-mass stars
($M<8$~$M_\odot$) we use the yields by Gavil\'an et
al.~(\cite{gavilan05}) for both values of the metallicity. We have
also tried the yields of Marigo (\cite{marigo01}), which however do
not give any appreciable difference in the computed gradients  of chemical
elements produced by intermediate mass stars, as N. For stars in the
mass range $8~M_\odot < M < 35~M_\odot$ we adopt the yields by Chieffi
\& Limongi~(\cite{chieffi04}) for $Z=0.006$ and $Z=0.02$. The yields
of more massive stars are affected by considerable uncertainties
associated to different assumptions about the modeling of processes
like convection, semiconvection, overshooting, mass loss. Other
difficulties arise from the simulation of the supernova explosion and
the possible fallback after the explosion, that strongly influences the
production of iron-peak elements. It is not surprising then that the
results of different authors (e.g. Arnett~\cite{arnett95}; Woosley \&
Weaver~\cite{woosley95}; Thielemann et al.~\cite{thielemann96}; Aubert
et al.~\cite{aubert96}) disagree in some cases by orders of magnitude
\footnote{Recently, Hirschi et al.~(\cite{hirschi05}) computed stellar
yields for massive stars of solar metallicity considering also the
effects of stellar rotation ($v_{\rm rot}=0$~km~s$^{-1}$ $v_{\rm
rot}=300$~km~s$^{-1}$), but only for a few chemical elements (not
including S).}.  In our models, we estimate the yields of stars in the
mass range $35~M_\odot <M<100~M_\odot$ by linear extrapolation of the
yields in the mass range $8~M_\odot < M < 35~M_\odot$.

\subsection{The IMF}
\label{sec_imf}

Recent work supports the idea that the IMF is universal in space and
constant in time (Wyse~\cite{wyse97}; Scalo~\cite{scalo98};
Kroupa~\cite{kroupa02}), apart from local fluctuations. As reviewed by
Romano et al.~(\cite{romano05}) there are several parameterizations of
the IMF that have been used in GCE models, starting from the original
Salpeter~(\cite{salpeter55}) power-law: piecewise power-laws
(Tinsley~\cite{tinsley80}; Scalo~\cite{scalo86}; Kroupa et
al.~\cite{kroupa93}; Scalo~\cite{scalo98}), polynomial approximations
(Ferrini et al.~\cite{ferrini90}), and logarithmic polynomials
(Chabrier~\cite{chabrier03}). Romano et al. (\cite{romano05}) studied
the sensitivity of GCE models to different parameterizations of the IMF
and found that the Scalo~(\cite{scalo86}), Kroupa et
al.~(\cite{kroupa93}) and Chabrier~(\cite{chabrier03}) IMFs are
generally more consistent with observational data than those of
Salpeter~(\cite{salpeter55}) and Scalo~(\cite{scalo98}).

One of the results most sensitive to the choice of the IMF is the
metallicity gradient in the disk. In fact, the magnitude and the slope
of chemical abundance gradients are related to the number of stars in
each mass range, and so to the IMF.  With our set of chemical yields
(see Sect.~\ref{sec_cheyields}), we find the best agreement with the
observed gradients adopting a two power-law IMF analogous to Kroupa's
IMF (with a slope $-2.35$ for $M>1$~$M_\odot$ and $-1.2$ for
$M<1$~$M_\odot$). This is the IMF adopted in this work.  Other
parameterizations, with the exception of Chabrier's IMF,  result in
general in an under-production of massive stars, and, consequently, a
lower production of metals.

We also examine the effects of a possible dependence of the slope of
the IMF on galactocentric radius. On the basis of HST photometric
observations of stellar clusters embedded in several giant H{\sc ii}
regions of M33, Lee et al.~(\cite{lee02}) found that the IMF becomes
steeper (less rich in massive stars) with increasing galactocentric
radius and decreasing metal abundance.  Assuming a variable slope in
the range $10~M_\odot<M_{\odot}<100~M_\odot$ as suggested by Lee et
al.~(\cite{lee02}), we find however that the resulting metallicity
gradient is not consistent with the observations, being characterized
by a much shorter scale length than the observed distribution of
metals.

\section{Results of the model}
\label{sect_results}

\begin{table*}
\centering   
  \begin{tabular}{lllllllll}
\hline
            & $R_D$         & $R_H$ & $\lambda_D$ &  $H (R_{\odot,{\rm M33}}$)       & $H^\prime (R_{\odot,{\rm M33}}$) 
  & $\mu (R_{\odot,{\rm M33}}$)                  & $f_0$               & ref. \\
            & (kpc)         & (kpc) & (kpc)       &(10$^{7}$~ yr)$^{-1}$& (10$^{7}$~yr)$^{-1}$&  (10$^{7}$~ yr) $^{-1}$ & (10$^{7}$~ yr)$^{-1}$    &       \\
\hline
M33          & 0.2    & 20  & 2   & 0.2    & 0.2   & 0.01    &  0.003 & {\em a}\\
\smallskip
M33          & 0.2    & 20  & 2   & 0.06   & 0.2   & 0.01    &  0.3 & {\em b}\\
\smallskip
Spiral $N=6$ &  --    & --  & --  & 0.012  & --    & 0.017   &  0.009  & {\em c} \\
\smallskip
M33          &  --    & --  & --  & 0.006  & --    & 0.006   &  0.0056  & {\em d}\\
\smallskip
MW           &  --    & --  & 2   & 0.5    & 1.0   & 0.05    &  0.7  & {\em e}\\
\hline
\end{tabular}
\caption{ Parameters of the GCE models for M33 and the MW.
$H$, $H^\prime$ and $\mu$ are given at the equivalent solar radius ($R_{\odot,{\rm M33}}$), 
$f_0$ is given at the centre. 
References: {\em a}, accretion (this work); {\em b}, collapse (this work); 
{\em c}, Moll\'a \& Diaz~(\cite{molla05}) for a spiral galaxy of morphological 
type $N=6$; {\em d}, Moll\'a et al.~(\cite{molla96}, \cite{molla97}), {\em e} Ferrini et al.~(\cite{ferrini92}) for the MW.}
\label{Tab_model}  
\end{table*}

In this section we present the results of our GCE model for M33
described in Sect.~\ref{sect_model}, and compare them to the data
discussed in Sect.~\ref{sect_data}. In particular, we examine the SFR,
the distribution of atomic and molecular gas, and the chemical
abundance gradients.

\subsection{The evolution of the SFR}

An important constraint to GCE models is the magnitude and the radial
profile of the SFR at the present epoch. This observable allows one to
discriminate between different evolutionary scenarios that result in
the same radial distributions of stars, gas and chemical abundances.

This is illustrated by the two models presented in
Fig.~\ref{Fig_infall} and in Table~\ref{Tab_model}, labeled {\em
collapse} and {\em accretion}, differing in the time behavior of the
accretion rate on the disk. In the former, the disk is formed by the
collapse of the gas initially present in the galaxy's halo, and the
resulting accretion rate of the disk decreases rapidly with time
(Fig.~\ref{Fig_infall}), simulating a rapid collapse phase.  Since the
exponential time scale of collapse is longer for the outer, less dense,
regions of the halo, the disk is formed in an inside-out fashion, as
clearly shown by the radial profile of the infall rate in
Fig.~\ref{Fig_infall}{\em b}. In the {\em accretion} model, the infall
rate is approximately constant for the entire evolution of the galaxy
(except the inner few kpc, where it decreases by about one order of
magnitude). In this case, the disk of M33 is built by continuous
gas supply from the external medium, a process  supporting a substantial
star formation  at all radii, in agreement with the observations
discussed in Sec.~\ref{sec_obs_infall}.

Both models well reproduce the  present time atomic and molecular
gas distributions, and the stellar density profile of M33.   The
differences between the two models are evident at earlier times,
because the bulk of the stellar mass settles in the disk earlier in the
{\em collapse} model than in the {\em accretion} model, due to the
rapid gas infall and to the high star formation rate. The behaviour of the
SFR reflects the differences in the two models at the present time:
the actual SFR predicted by the  {\em accretion} model is in closer
agreement with the observations than the SFR predicted  by the {\em
collapse} model. In the {\it collapse} model the accretion rate and SFR vary
more rapidly with time: as a consequence stars form earlier and the
present-day SFR is at least a factor $\sim 3$ lower than what can be inferred by
infrared and optical observations. Clearly, the higher infall rate on the disk 
at present times for the {\em accretion} model results in a more vigorous SFR than 
in the {\em collapse} model. But at earlier times the SFR was higher for
the {\em collapse} model, since the larger gas reservoir accreted earlier
by the disk was converted into stars.  The present-day SFR integrated over the disk
up to $R=10$~kpc is $0.2$~$M_\odot$~yr$^{-1}$ for the {\em collapse}
model and $0.5$~$M_\odot$~yr$^{-1}$ for the {\em accretion} model, 
but the difference increases  in the inner regions of the disk.  The integral of
the observed SFR up to $R=7$~kpc, $\sim 0.4$~$M_\odot$~yr$^{-1}$ (see
Sect.~\ref{sec_sfr}), is in good agreement with the {\em accretion}
model results. Also the intermediate-age SFR derived from PNe, 
$\sim 0.55$~$M_\odot$~yr$^{-1}$ from 1 to 4~Gyr ago, and
$\sim 1.1$~$M_\odot$~yr$^{-1}$ from 1 to 8~Gyr ago, are in closer agreement
with the {\em accretion} model which predicts $0.7$~$M_\odot$~yr$^{-1}$
and $0.9$~$M_\odot$~yr$^{-1}$ respectively.  In the rest of the paper we
adopt the {\em accretion} model to describe the evolution of M33.

\subsection{The radial distribution of gas and stars}

\begin{figure}
\centering
\resizebox{\hsize}{!}{\includegraphics[angle=-90]{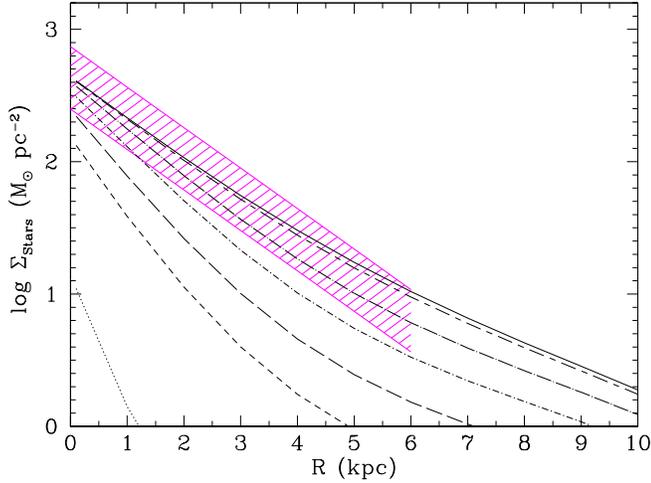}}
\caption{Time evolution of the stellar surface density (line
types as in Fig.~\ref{Fig_sfr}), compared with the stellar surface
density determined by Corbelli~(\cite{corbelli03}, {\em shaded region}).}
\label{Fig_stars}
\end{figure}

\begin{figure}
\centering
\resizebox{\hsize}{!}{\includegraphics[angle=-90]{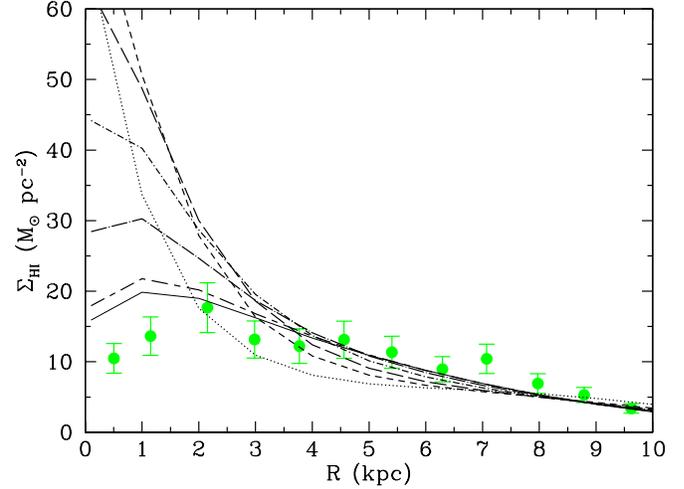}}
\caption{Time evolution of the surface density of diffuse gas (line
types as in Fig.~\ref{Fig_sfr}), compared with the observed surface
density of atomic hydrogen (Corbelli~\cite{corbelli03}).}
\label{Fig_hi}
\end{figure}

\begin{figure}
\centering
\resizebox{\hsize}{!}{\includegraphics[angle=-90]{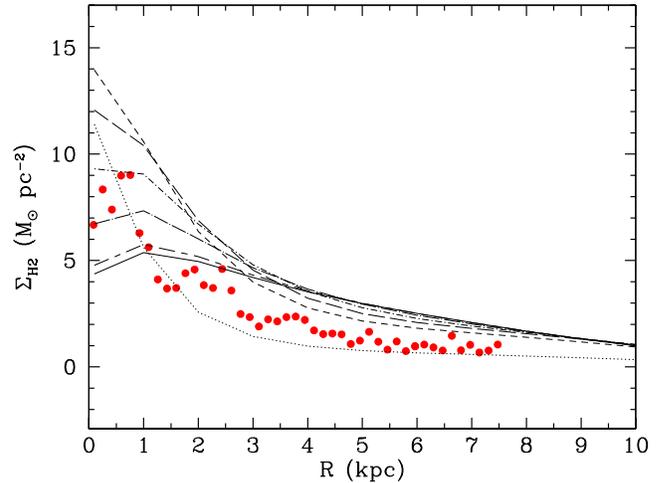}}
\caption{Time evolution of the surface density of clouds (line types as
in Fig.~\ref{Fig_sfr}), compared with the observed surface density of
molecular hydrogen (Corbelli~\cite{corbelli03}).}
\label{Fig_hii}
\end{figure}

We show in Figures~\ref{Fig_stars}, \ref{Fig_hi}, \ref{Fig_hii}, the
time evolution of the surface densities of stars, atomic, and molecular
gas.  The diffuse gas is accreted by the disk from the external medium
following an exponential law which produces a larger amount of gas in
the central regions of the galaxy.  Clouds form out of this diffuse
gas, whose surface density decreases with time.  Stars are formed at
the expense of clouds, and the cloud surface density decreases with
time while the stellar surface density increases.  The present-time
radial distribution of diffuse and condensed gas can be compared with
the observed radial profiles of atomic and molecular gas,
respectively.  The model predicts a slightly higher surface density of
diffuse gas in the central regions and a lower cloud surface density
than observed. This might either imply that the formation of diffuse
gas via clouds collisions is even more inefficient at small radii than
we assumed, or that there is an enhanced formation of molecular
clouds.  This might be due to additional features in place in the
central regions, such as a small bar fueling gas towards the center or
to a small bulge with a higher metal and dust abundance enhancing the
molecular hydrogen fraction.  The slightly lower abundance of clouds
seen at large galactic radii might be associated with a change of the
CO-H$_2$ conversion factor due to, e.g. a lower excitation rate of the
CO molecule.


\subsection{The metallicity gradients}
\label{sec_gradients}

We now examine the evolution of the abundance gradients of O, Ne, S, N,
and Fe in M33. In this nearby galaxy is possible to separate the
present-day metallicity gradient, outlined by abundances of H{\sc II}
regions, A-type and B-type supergiant, and Cepheids, from an older
metallicity gradient, outlined by PNe and RGB stars. The PNe trace the
chemical composition of the ISM over a range of galactic ages between
$\sim 1$ to $\sim 8$~Gyr ago, at least for those elements not affected
by stellar evolution in the mass range $M<$8~M$_\odot$, such as O, Ne
(only as a first approximation, cf.  Marigo~\cite{marigo01}; Magrini et
al.~\cite{m05}), and S. Abundances in RBG stars trace the metallicity
gradient relative to an even distant past ($>$8~Gyr ago).

In order to avoid statistical effects due to the incompleteness of
the various samples, in the rest of this Section we shall compare the
model results with larger and homogeneous samples of chemical abundance
determinations for each class of objects, namely H{\sc ii} regions,
young stars, and RGB stars.  As discussed by e.g Vilchez et
al.~(\cite{vilchez88}), Urbaneja et al.~(\cite{urbaneja05}), Magrini et
al.~(\cite{m07}), the gradient in M33 does not have not a constant slope, 
and it cannot be represented by a single power-law: studies undertaken for
different radial ranges would produce different results.  Inclusion of 
the innermost regions would result in a steeper gradient. On the contrary
the gradient flattens out if only regions far from the center were considered. 
Thus, for each chemical element we compute  the gradient
considering  all data from the literature within a given radial range,
and we include also new abundance determinations by  Magrini et
al.~(\cite{m07}).  In Table~\ref{table:grad} we list the slope of the
gradients and the references for the data used to derive it.

\begin{table*}
\label{table:grad}      
\begin{tabular}{llllllllllll}        
\hline
\smallskip
    &  Model       &           &           & Observations      &              &                & \multicolumn{2}{c}{References}&              &                &      \\
\hline
\smallskip
    & Present  & 5~Gyr ago & 8~Gyr ago & H{\sc ii} regions & young stars  & PNe            & RGB    & H{\sc ii} regions  & young stars  & PNe            & RGB  \\ 
    &              &           &           &\multicolumn{2}{c} {present time}      & 1-8~Gyr ago    &$>8$~Gyr &              &              &                &\\
\hline
\smallskip
     &            &           &           &                 &                 &                 &           &             &              &                &\\
O/H  & $-0.067$   & $-0.078$  & $-0.094$  & $-0.07\pm 0.01$ & $-0.07\pm 0.02$ & $-0.11\pm 0.05$ &           & {\em a}, {\em b}, {\em c}  & {\em g}, {\em h}, {\em k}  & {\em j}      & \\
     &            &           &           &                 &                 &                 &           & {\em d}, {\em e}         &           &                 &  \\
\smallskip
Ne/H & $-0.067$   & $-0.077$  & $-0.091$  & $-0.06\pm 0.02$ &                 &                 &           & {\em f}, {\em d}           &              &  {\em j}              &\\
     &            &           &           &                 &                 &                 &           &              &              &                &\\
\smallskip
S/H  & $-0.072$   & $-0.082$  & $-0.095$  & $-0.07\pm 0.05$ &                 & $-0.09\pm 0.04$ &           & {\em b}, {\em c}, {\em e}         &              &                &\\ 
     &            &           &           &                 &                 &                 &           &              &              &                &\\
\smallskip
N/H  & $-0.125$   & $-0.137$  & $-0.149$  & $-0.10\pm 0.02$ &                 &                 &           &{\em a}, {\em b}, {\em c}, {\em e}      &              &                &\\ 
     &            &           &           &                 &                 &                 &           &              &              &                &\\
\smallskip
Fe/H & $-0.062$   & $-0.069$  & $-0.080$  &                 &                 &                 & $-0.07\pm 0.01$ &        &              &                & {\em i}, {\em l}, {\em m} \\     
&                 &           &           &                 &                 &                 &           &              &              &                & {\em n}, {\em o}, {\em p} \\
\hline
\end{tabular}
\caption{ Metallicity gradients (dex~kpc$^{-1}$) from 1 to 10~kpc.  
The observed
gradients are computed from the data discussed in Sect.~\ref{sect_data_chem} 
using a weighted linear fit.  
Model gradients  are approximated by linear fits.
References: 
{\em a}, Smith~(\cite{smith75}); 
{\em b}, Kwitter \& Aller~(\cite{kwitter81}); 
{\em c}, Vilchez et al.~(\cite{vilchez88}); 
{\em d}, Crockett et al.~(\cite{crockett06});
{\em e}, Magrini et al.~(\cite{m07});
{\em f}, Willner \& Nelson-Patel~(\cite{willner02}); 
{\em g}, Monteverde et al.~(\cite{monteverde97}, \cite{monteverde00}); 
{\em h}, Urbaneja et al.~(\cite{urbaneja05}); 
{\em k}, Beaulieu et al.~(\cite{beaulieu06}); 
{\em j}, Magrini et al.~(\cite{m04}); 
{\em i}, Barker et al.~(\cite{barker06}); 
{\em l}, Stephens \& Frogel~(\cite{stephens02}); 
{\em m}, Kim et al.~(\cite{kim02}); 
{\em n},  Galletti et al.~(\cite{galletti04}); 
{\em o}, Tiede et al.~(\cite{tiede04}); 
{\em p}, Brooks et al.~(\cite{brooks04}).}
\end{table*}

\subsubsection{The abundance of oxygen}

\begin{figure}
\resizebox{\hsize}{!}{\includegraphics[angle=0]{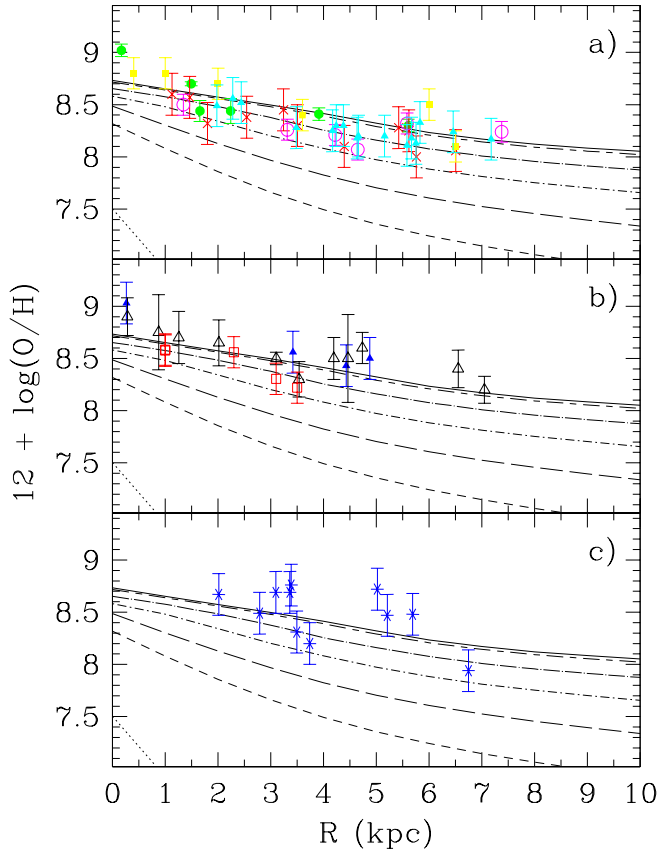}}
\caption{ The time evolution of the O gradient (line types
as in Fig.~\ref{Fig_sfr}) compared with: a) H{\sc ii} region
determinations:
Smith~(\cite{smith75}, {\em filled squares}\/); 
Kwitter \& Aller~(\cite{kwitter81}, {\em crosses}\/);
Vilchez et al.~(\cite{vilchez88}, {\em filled circles}\/);
Crockett et al.~(\cite{crockett06}, {\em empty circles}\/);
Magrini et al.~(\cite{m07}, {\em filled triangles}\/); 
b) B supergiant stars: 
Monteverde et al.~(\cite{monteverde97}, \cite{monteverde00}, {\em filled triangles}\/); 
Urbaneja et al.~(\cite{urbaneja05}, {\em empty triangles}\/); 
Cepheids: Beaulieu et al.~(\cite{beaulieu06}, {\em empty squares}\/); 
c) PNe abundances: 
Magrini et al.~(\cite{m04}, {\em crosses}\/).}
\label{Fig_oxy}
\end{figure}

In Fig.~\ref{Fig_oxy} we compare the time evolution of the O gradient
with observations of H{\sc ii} regions ($a$), stars ($b$) and PNe
($c$).  The O/H gradient predicted by the model at present time is
$-0.067$~dex~kpc$^{-1}$ from 1 to 10~kpc in radius.  This is in
agreement with the gradient derived in the same radial range from 
a combined sample composed by abundances in H{\sc ii} regions
available in the literature and from recently acquired data presented
in Magrini et al.~(\cite{m07}).  A linear fit of all H~{\sc ii}
regions data, which references are quoted in Table~\ref{table:grad},
gives $-0.07\pm 0.01$~dex~kpc$^{-1}$ over a range of radii from 1 to 10~kpc,
even though the data show that the gradient flattens out going radially
outwards.  We stress the need to assemble a statistically significant
sample in order to avoid the influence of intrinsic peculiarities of
the observed sources and inhomogeneities associated to particular
regions e.g. to spiral arms.

The correspondence between the predicted O/H gradient and that
determined from absorption lines in young stars is also very good.
Monteverde et al.~(\cite{monteverde00}) find $-0.078 \pm
0.06$~dex~kpc$^{-1}$, and a similar value was also found by Urbaneja et
al.~(\cite{urbaneja05}), $-0.06 \pm 0.02$~dex~kpc$^{-1}$.  The overall
linear fit to O abundances from young stars, including A, B giant stars
and Cepheids, (references quoted in Table~\ref{table:grad}),
gives $-0.07\pm0.02$~dex~kpc$^{-1}$ (Pearson correlation factor
$-0.6$), in agreement with the model result (see
Table~\ref{table:grad}).

The O/H gradient of a sample of PNe  reaches a lower absolute
value of abundances at large radii and  is steeper than the O/H gradient
outlined by H{\sc ii} regions and young stars. A weighted linear fit
gives a slope $-0.10 \pm 0.05$~dex~kpc$^{-1}$ (with Pearson
correlation factor $-0.7$). This behaviour is expected in a scenario
where metallicity gradients flatten with time  because chemical 
abundances at large radii increase gradually with time while the
enrichment process is very fast in the central regions. The O/H gradient from 
PNe is representative of the metallicity in the disk of M33 about 1--8~Gyr
ago. The same is shown by our model (compare the long-short dashed
curve and the short dash-dotted line in Fig.\ref{Fig_oxy}$c$). More
quantitatively, the slopes of the O/H gradients predicted by our model
are $-0.078$~dex~kpc$^{-1}$ and $-0.094$~dex~kpc$^{-1}$, 5 and 8~Gyr
ago respectively (see Table~\ref{table:grad}).

Our sample includes some of the brightest PNe of M33, probably representative
of a younger population than the average sample (Richer et
al.~\cite{richer98}). Five PNe show an O abundance in agreement with the model
predictions, four are marginally consistent, and two are over-abundant
in O (PN~75 and 91, see Magrini et al.~\cite{m04}).  A possible explanation 
for the O over-abundance in some PNe is
the occurrence of hot bottom burning and third dredge-up processes
during the post-AGB phases, with a consequent O enrichment in the
nebula (Marigo~\cite{marigo01}, Herwig~\cite{herwig04}).  These
processes are generally associated with massive progenitors and
therefore with N enrichment as well.  We have therefore excluded 
the two PNe over-abundant in O from the weighted
linear fit shown in Table~\ref{table:grad}.

\subsubsection{The abundance of neon}

\begin{figure}
\resizebox{\hsize}{!}{\includegraphics[angle=-90]{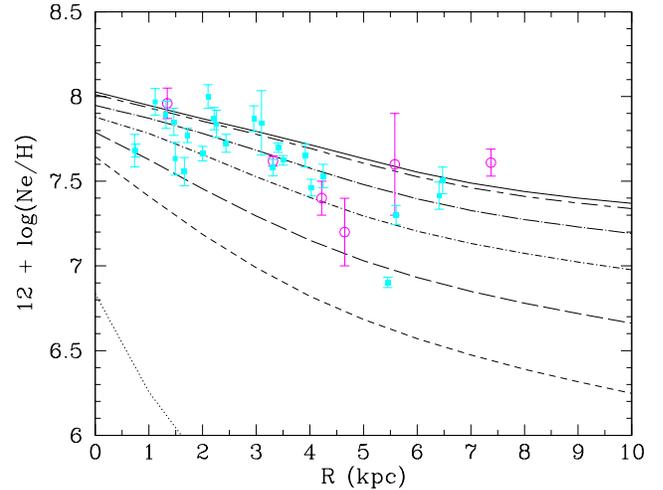}}
\caption{The time evolution of the Ne gradient (line types
as in Fig.\ref{Fig_sfr}). Comparison with the observations:
H{\sc ii} regions infrared spectra, Willner \& Nelson-Patel~(\cite{willner02}, 
{\em filled squares}) and optical spectra, 
Crockett et al.~(\cite{crockett06}, {\em empty circles}).}
\label{Fig_neon}
\end{figure}

O and Ne are both produced mainly by short-lived, massive stars ($M >
10$~$M_\odot$). Stellar nucleosynthesis models (e.g. Chieffi \&
Limongi~\cite{chieffi04}, Hirschi et al.~\cite{hirschi05}), predict
that the abundances of these two elements should be closely correlated.
This prediction is supported by abundance measurements in Galactic and
extragalactic PNe (Henry~\cite{henry90}), showing that Ne/O is constant
over a wide range of [O/H] values.  Fig.\ref{Fig_oxy} and
\ref{Fig_neon}, show indeed that the slopes of the two gradients
predicted by our model are indistinguishable and equal to
$-0.067$~dex~kpc$^{-1}$.

Willner \& Nelson-Patel~(\cite{willner02}) derived the Ne abundance for 25
H{\sc ii} regions in M33 from infrared spectroscopy. They find an
inconsistency with the O gradient measured by Vilchez et
al.~(\cite{vilchez88}, $-0.12 \pm 0.01$~dex~kpc$^{-1}$.  Excluding the
inner two and the outer three regions of their sample they found a
best-fit slope of $-0.07 \pm 0.02$~dex~kpc$^{-1}$, in agreement with
our predictions. An even shallower slope for the [Ne/H] gradient is
found by Crockett et al.~(\cite{crockett06}), $-0.016$~dex~kpc$^{-1}$.
Both sets of observations are plotted in Fig.\ref{Fig_neon} for a
comparison with our predicted gradient.
Although there is some dispersion in the data in the central and outer
regions, the overall slope of the observed Ne/H gradient is $-0.06\pm
0.02$~dex~kpc$^{-1}$ (with Pearson correlation factor $-0.7$).  This
has been computed using data available in references given in 
Table~\ref{table:grad}, and is in agreement with the gradient  
of $-0.067$~dex~kpc$^{-1}$ predicted by our model.

\subsubsection{The abundance of sulfur}

The abundance of S in PNe is expected to be a better tracer of the
chemical composition of the ISM at the time of formation of PNe
progenitors than the abundance of O and Ne (Henry et
al.~\cite{henry04}). For this reason, the S/H gradient in PNe has been
used in our Galaxy to determine the temporal behaviour of the
metallicity gradient (Maciel et al. \cite{maciel05}, \cite{maciel06}).
We recall however the possible sulfur ``anomaly'' seen in some Galactic
PNe (Henry et al.~\cite{henry04}):  when compared with similar data for
stars and H{\sc ii} regions, some Galactic PNe have much lower S
abundance than expected on the basis of their O abundance.  This
anomaly is not dominant in the sample of PNe by Magrini et
al.~(\cite{m04}).  The average $\log({\rm S/O})=-1.92$ is consistent
with Galactic studies of both PNe and H{\sc ii} regions (cf. $-1.91 \pm
0.24$, Henry et al.~\cite{henry04}). It might be present at most in two
PNe (PN93 and PN96) with very low S/O ratio.

\begin{figure}
\resizebox{\hsize}{!}{\includegraphics[angle=0]{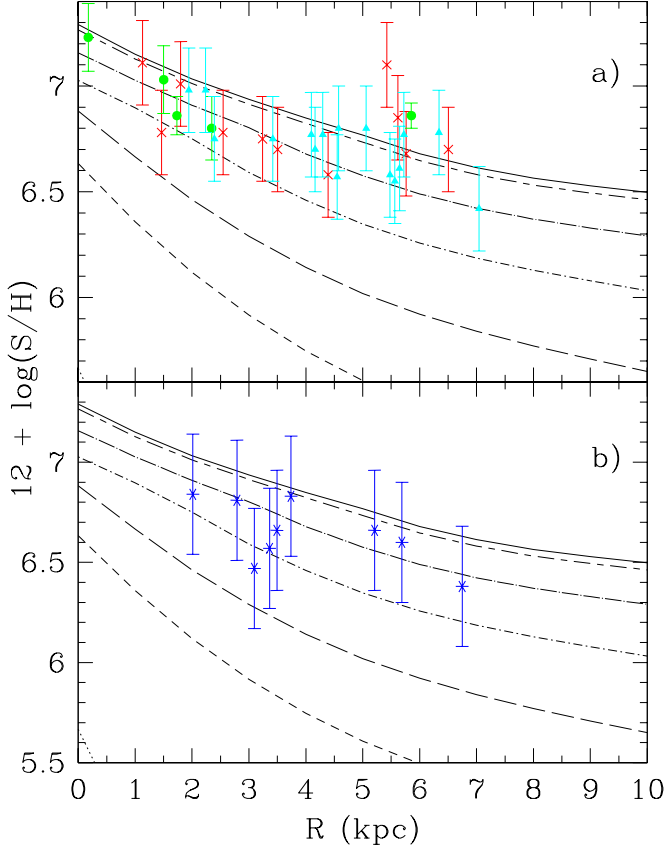}}
\caption{The time evolution of the S gradients (line types as in
Fig.\ref{Fig_sfr}) compared with the observations of: a) H{\sc ii}
regions: 
Kwitter \& Aller~(\cite{kwitter81}, {\em crosses}\/); 
Vilchez et al.~(\cite{vilchez88}, {\em filled circles}\/);
Magrini et al.~(\cite{m07}, {\em filled triangles}\/); 
b) PNe: 
Magrini et al.~(\cite{m04}, {\em crosses}\/).}
\label{Fig_sulph}
\end{figure}

In Fig.\ref{Fig_sulph}{\em a} and {\em b} we show the S abundance of
H{\sc ii} regions and PNe, respectively. Notice the good agreement of
the present-day S gradient predicted by our model
($-0.07$~dex~kpc$^{-1}$), with the S abundances gradient for H{\sc ii}
regions ($-0.07 \pm 0.05$~dex~kpc$^{-1}$, Pearson correlation factor
of $-0.6$). A best-fit to the PN sample gives an S gradient of $-0.09
\pm 0.04$~dex~kpc$^{-1}$ (with a Pearson correlation factor $-0.8$)
consistent with the value $-0.082$~dex~kpc$^{-1}$ predicted by our
model 5~Gyr ago.  We have excluded the two PNe possibly affected by
the sulfur ``anomaly''.  These do not correspond to the two PNe
excluded from the oxygen gradient, since the two phenomena, sulfur ``anomaly'' 
and oxygen overabundance, are not necessarily linked.

\subsubsection{The abundance of nitrogen}

\begin{figure}
\centering
\resizebox{\hsize}{!}{\includegraphics[angle=-90]{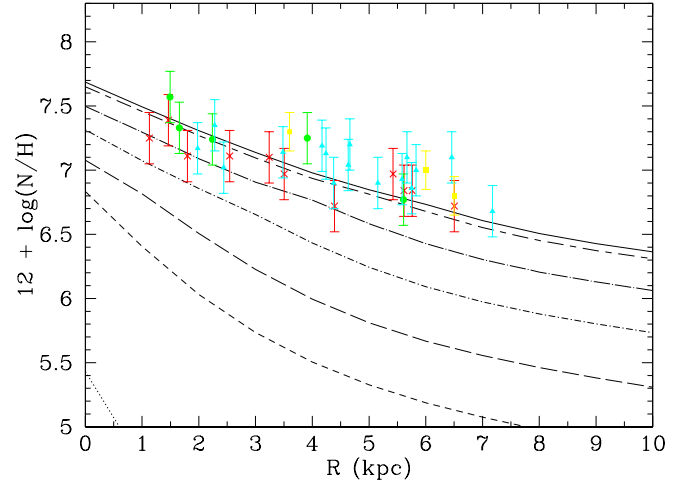}}
\caption{The time evolution of the N/H gradient according to our model
(line types as in Fig.~\ref{Fig_sfr}), compared with measurements of
N in H{\sc ii} regions: Smith~(\cite{smith75}, {\em filled squares}\/);
Kwitter \& Aller~(\cite{kwitter81}, {\em crosses}\/);
Vilchez et al.~(\cite{vilchez88}, {\em filled circles}\/);
Magrini et al.~(\cite{m07}, {\em triangles}\/).}
\label{Fig_nitr}
\end{figure}

Unlike O, Ne and S, nitrogen is mainly produced by intermediate mass
stars. Therefore, the N abundance of B stars (Urbaneja et
al.~\cite{urbaneja05}) and PNe (Magrini et al.~\cite{m04}) is affected
by nucleosynthesis processes, and cannot be used to infer the original
chemical composition of the ISM.  In Fig.\ref{Fig_nitr} we show the
model evolution of the N/H gradient and the abundances observed in
H{\sc ii} regions.  The predicted slope of the present-day N gradient,
$-0.125$~dex~kpc$^{-1}$, is in good agreement with H{\sc ii} regions
data ($-0.10\pm 0.02$~dex~kpc$^{-1}$, with a Pearson correlation factor
$-0.8$).

The general agreement of the observed gradients of chemical elements
produced by stars in different mass ranges with the model supports the
reliability of the adopted IMF.  Any choice of the IMF that predicts
the correct number of massive stars, but not of low- and intermediate
mass stars, would reproduce equally well the O/H, S/H and Ne/H
gradients, but would fail in predicting the N radial distribution.

\subsubsection{The abundance of iron}
\label{sec_iron}

\begin{figure}
\centering
\resizebox{\hsize}{!}{\includegraphics[angle=-90]{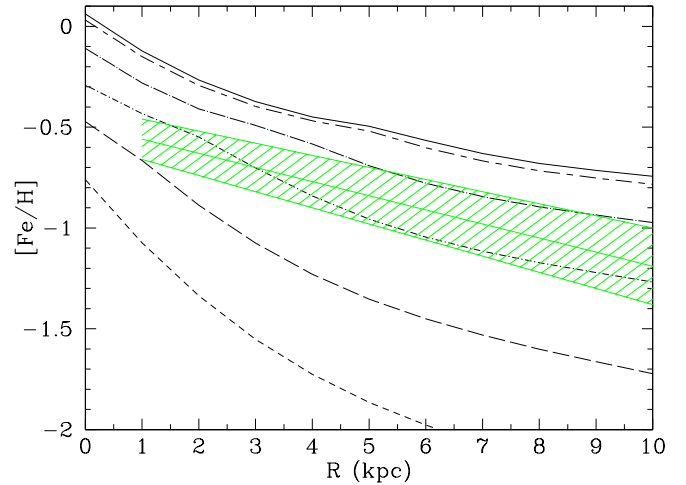}}
\caption{The time evolution of the [Fe/H] gradient according to our model
(line types as in Fig.\ref{Fig_sfr}), compared with the [Fe/H] gradient
of RGB stars (Barker  et al.~\cite{barker06}, {\em shaded area}\/).}
\label{Fig_iron}
\end{figure}

Barker et al.~(\cite{barker06}) analyzed [Fe/H] abundances\footnote{ As
usual, $[\rm{Fe/H}]=\log(\rm{Fe/H})-\log(\rm{Fe/H})_\odot$ and
$12+\log(\rm{Fe/H})_\odot=7.45$ (Asplund et al.~\cite{asplund05}).} in
RGB stars of M33  from new observations and from  previous works by
Stephens \& Frogel~(\cite{stephens02}), Kim et al. (\cite{kim02}),
Galletti et al.~(\cite{galletti04}), Tiede et al.~(\cite{tiede04}),
Brooks et al.~(\cite{brooks04}).  Using these data, they found a
well defined gradient $[{\rm  Fe/H}]=-0.07R-0.49$ extending up to
$R\approx 12$~kpc.  Fig.\ref{Fig_iron}, shows the good agreement
between the time evolution of Fe/H predicted by our model and the RGB
abundances of Barker et al.~(\cite{barker06}). The epoch of formation
of RGB stars is $>$8~Gyr ago, when the slope of the Fe/H gradient
according to our model is $-0.08$~dex~kpc$^{-1}$.  Note however that
the linear fit is a very rough approximation to the effective model
gradient.

The metallicity gradient of Cepheids determined by Beaulieu et
al.~\cite{beaulieu06}, -0.2~dex~kpc$^{-1}$, can also be compared with
the predicted gradient at present time. However, the five observed Cepheids
are located in the radial region from $\sim$1 to 4~kpc,
whereas our computed gradient extends out to 10~kpc.  In the inner
region of M33, the [Fe/H] gradient predicted by our {\em accretion} and
{\em collapse} models model are $-0.11$~dex~kpc$^{-1}$ and
$-0.14$~dex~kpc$^{-1}$, respectively. We must keep in mind however that
the gradient quoted by Beaulieu et al.~(\cite{beaulieu06}) refers to
the total metallicity, not to the Fe/H gradient.

\subsection{Comparison with previous models}

 Specific models for the chemical evolution of M33 have been
previously developed by Diaz \& Tosi~(\cite{diaz84}), Moll\'a et
al.~(\cite{molla96}, \cite{molla97}), and by Moll\'a \&
Diaz~(\cite{molla05}). Our approach is very similar to that of Moll\'a et
al.~(\cite{molla96}, \cite{molla97}), and the predicted H~I and H$_2$ radial
distributions are indeed similar. The chemical
gradients predicted by the model of Moll\'a et
al.~(\cite{molla96}, \cite{molla97}) are however steeper 
(e.g. d(O/H)/dR $\sim$ -0.21 dex/kpc, d(N/H)/dR $\sim$
-0.34 dex/kpc) than what the most recent abundance
determinations suggest.  This is probably a consequence of the assumed IMF,
heavily weighted towards massive stars in the parametrization adopted
by Moll\'a et al.~(\cite{molla96}, \cite{molla97}).  The time
evolution of the metallicity gradients discussed in detail in Moll\'a
et al.~(\cite{molla97}), shows the same behaviour as in the present
work, i.e. a flattening with time.  Metallicity gradients resulting
from the most recent models by Moll\'a
\& Diaz~(\cite{molla05}), not computed specifically for M~33 but
applied to the M~33 case, are also steeper than what the collection of
observations presented here seem to show. Moll\'a
\& Diaz~(\cite{molla05}) assume a slightly higher infall rate than 
Moll\'a et al.~(\cite{molla96}, \cite{molla97}), 
resulting in significantly higher surface density of H~I and H$_2$,
which are marginally consistent with the data. In particular their model 
predicts a strong decrease of the H~I surface density towards the center, a
large central H~I hole, and a slight decrease of the H$_2$ surface
density in the same region, which are not observed in this galaxy.

\section{The time evolution of metallicity gradients}
\label{sect_grad}

Both models we consider predict abundance gradients which flatten with time. 
The main difference between the two model is the resulting value of
the slope of the gradients, that are much steeper at all times in the
{\em collapse} model than in the {\em accretion} model. This is due to
the different nature of the infall: the {\em collapse} model predicts
a much shorter time scale for the formation of the disk,  due to a
rapid collapse of the halo, while the {\em accretion} model predicts a
continuous infall of material from the intergalactic medium.

In the case of M33, the best fit to the whole set of  observations is
obtained with our {\em accretion} model, or, in the terminology
by Moll\'a et al.~\cite{molla93}, with a model which has a self consistent  
SFR and a continuous gas infall.
This model does not exclude an early collapse phase of a galactic halo,
but considers its effects less important at the present time than the
accumulation of matter from the intergalactic medium.  This is
supported by recent observations (see Sect.~\ref{sec_obs_infall}) and
by numerical simulation of galaxy formation. As discusses in
Sect.~\ref{sec_infall} the intergalactic supply is important,
especially in low mass halos, and, in the case of M33. The continuous
supply of gas is necessary to reproduce the high SFR observed at
present time (see Fig.~\ref{Fig_sfr}).

\begin{figure}
\centering
\resizebox{\hsize}{!}{\includegraphics[angle=0]{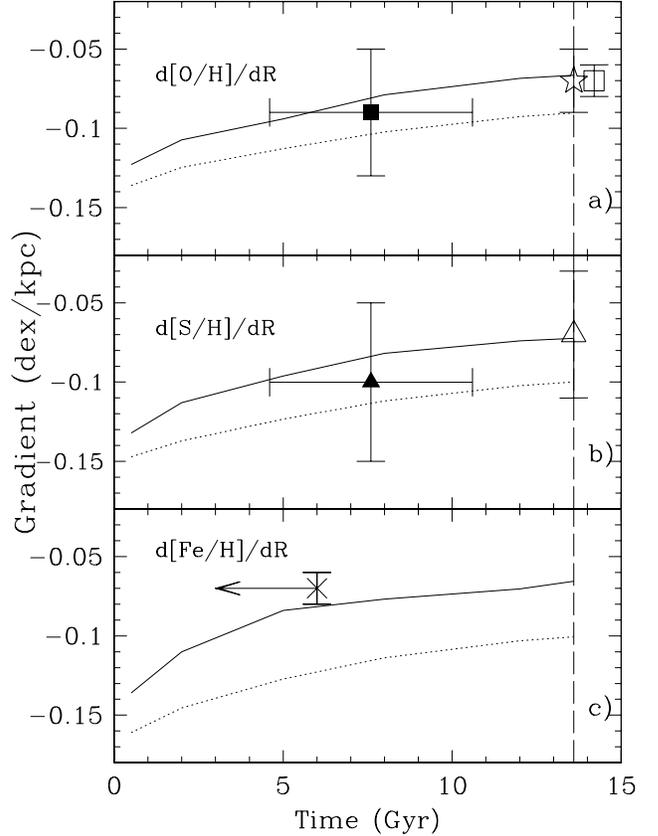}}
\caption{The time evolution of the [O/H] (a), [S/H] (b), and [Fe/H](c)
gradients according to our {\em accretion} (continuous line) and {\em
collapse} (dotted line) models, compared with gradient slopes of sulfur
({\em filled triangle}) and oxygen ({\em filled square}) of  PNe, of
sulfur ({\em empty triangle}) and oxygen ({\em empty square}) of H{\sc
ii} regions and  of young giant stars ({\em star}), and of iron of RGB
stars ({\em cross}). The vertical dashed line marks the present age of
the galaxy, 13.6 Gyr.}
\label{Fig_gradev}
\end{figure}

In Fig.~\ref{Fig_gradev} we compare the time evolution of the slope of
the gradient of O/H and S/H (panel {\em a}, {\em accretion} model;
panel {\em b}, {\em collapse} model) with the gradient measured in
young stars, H{\sc ii} regions and PNe. To derive the gradient, both
the model predictions and the observational data have been linearly
fitted in the $\log[M/H]$--$\log R$ plane, in the range of galactic
radii 1--10~kpc.  The flattening of the metallicity gradient predicted
by the {\em accretion} model over the entire lifetime of the disk of
M33 is quite modest because of the slow process of formation of the
disk, which is still taking place at the present time. On the other
hand, the {\em collapse} model predicts steeper gradients at all times,
especially in the past, because of the rapid initial collapse phase
that forms the disk in a few Gyr.  This behaviour is in conflict with
the observations, especially with the Fe/H gradient of RGB stars (see
Sect.~\ref{sec_iron}) determined by Barker et al~(\cite{barker06}), as
shown in Fig.~\ref{Fig_gradev}{\em c}.

The comparison of the abundances in PNe with that in H{\sc ii} regions
presented in Sect.~\ref{sec_gradients}, represents the first evidence
for a flattening of the abundance gradients with time in an external
galaxy. A similar result has been found in the MW by Maciel et
al.~(\cite{maciel06}). The flattening in M33 is particularly evident
for O and S abundances, as shown by Fig.~\ref{Fig_oxy} and
Fig.~\ref{Fig_sulph}. Another feature of particular interest in the
case of M33 is the overall increase of metallicity with time at
all radii. The steady infall rate, that increases the gas content of
the disk with time, fuels star formation at all radii. The flattening
of the gradient with time is due to the combination of the weak radial
dependence of the infall rate and to the star formation efficiency
which decreases more rapidly with time in the central regions.

\section{Conclusions}
\label{sect_conc}

We have analyzed a large sample of observations for the LG spiral
galaxy M33, including  gas and  stellar radial distributions,
metallicity gradients, and  star formation rate, and we have
interpreted them using GCE models for this galaxy. We summarize here
our main results:

\begin{itemize}
\item[({\em i}\/)] Our {\em accretion} model (where the infall rate on
the disk is almost constant in time) reproduces the data better than
our {\em collapse} model (where the accretion rate decreases rapidly to
simulate the early collapse  phase of a baryonic halo).  Whereas
both models can account (within the errors) for the observed
distribution of stars, atomic and molecular gas,  the abundance 
gradients, outlined by a large number of abundance
determinations, and the high value of the present-day SFR allow one to
discriminate between different accretion and SFR histories for this
galaxy.

\item[({\em ii}\/)] The ability of our {\em accretion} model to
reproduce the observed constraints suggests the existence of an
extended phase for the formation of the disk of M33.  
The continuous accretion of material from the
intergalactic medium, is also supported by deep observations at
21-cm in the proximity of M33 and by numerical simulation of galaxy
formation and evolution. The continuous infall of gas into the disk is
necessary to maintain a high SFR, declining very slowly with time.

\item[({\em iii}\/)] Chemical abundances determined in young stars,
H{\sc ii} regions, PNe, and RBG stars, all indicate a relatively flat
radial distribution of elements, suggesting a slow, continuous rate of
star formation in the disk of M33. The magnitude and evolution of the
gradients of O/H, Ne/H, N/H, S/H, [Fe/H] computed with our {\em
accretion} model are in good agreement with these observations.  In
particular, S/H and O/H gradients of PNe when compared with the same
elemental abundance gradients of H{\sc II}, and the [Fe/H] gradient of
RGB stars, give independent observational evidence that the metallicity
gradient has flattened over the last $\sim 8$~Gyr, in agreement with
our model results.  A similar behaviour has been recently found in the
MW (Maciel et al.~\cite{maciel06}) and may represent a general feature
of the evolution of disk galaxies, even though galaxies of different
masses, such as the MW and M33, might have followed different
evolutionary histories.

\end{itemize}

\begin{acknowledgements}
We thank  the referee M. Moll\'a for her careful reading of the manuscript
and  for the detailed suggestions/comments that improved considerably 
the quality of the work presented in this paper. 
We are grateful to F. Ferrini for making his chemical evolution code 
available to us, and R. Walterbos for providing the H$\alpha$ map of M33.
The work of LM is supported by a INAF post-doctoral grant 2005.
\end{acknowledgements}

\begin{appendix}
\section{Chemical abundances in M33}
We present here the collection of chemical abundance measurements used in the present work.

\end{appendix}
\scriptsize{
\longtab{1}{
\begin{longtable}{rllllllll}
\caption{Chemical abundances used to constrain the present chemical evolution model of  M~33 from {\sc H~II} regions (S75, Smith~\cite{smith75}; 
KW81, Kwitter \& Aller \cite{kwitter81}; V88, Vilchez et al.~\cite{vilchez88}; C06,  Crockett et al.~\cite{crockett06}; 
M07, Magrini et al.~\cite{m07}), young stars (M97, Monteverde et al.~\cite{monteverde97}; U05, 
Urbaneja et al.~\cite{urbaneja05}; B06, Beaulieu et al.~\cite{beaulieu06}),  PNe (M04, Magrini et al.~\cite{m04}).
Chemical abundance marked with * are from  recompilation by Garnett et al.~(\cite{garnett97}) and those marked with : are 
derived without the electron temperature measurement.
We adopted typical errors of 0.2 dex when they are not quoted.
 }\label{Tab_abund_chim}\\
\hline\hline
Type & Name     & RA        & Dec             & O/H & N/H & Ne/H & S/H & Ref.\\
     &          & \multicolumn{2}{c}{J2000.0} &     &     &      &     &         \\
\hline
\endfirsthead
\caption{continued.}\\
\hline\hline
Type & Name     & RA        & Dec             & O/H & N/H & Ne/H & S/H & Ref.\\
     &          & \multicolumn{2}{c}{J2000.0} &     &     &      &     &         \\
\hline
\endhead
\hline
\endfoot
H~II & BCLMP 93 (CC93)&  1 33 52.3  & +30 39 18        & 8.85*    & -       & -     &  -   &  S75 :\\
     & BCLMP 87 (CC87)&  1 34 01.0  & +30 38 56        & 8.80*    & -       & -     &  -   &  S75 :\\
     & NGC 595        &  1 33 35.50 & +30 41 52.0      & 8.70*    & -       & -     &  -   &  S75 :\\
     & NGC 604        &  1 34 33.19 & +30 47 05.6      & 8.35*    & 7.30*   &  -    &  -   &  S75    \\
     & NGC 588        &  1 32 45.2  & +30 38 54.0      & 8.45*    & 7.00*   &  -    &  -   &  S75    \\
     & IC 132         &  1 33 15.8  & +30 56 45.0      & 8.05*    & 6.75*   &  -    &  -   &  S75    \\
\hline
     & BCLMP 85 (MA11)&  1 34 07.0  & +30 39 23        & 8.70*   & 7.35*   & -     & 7.11 & KW81 : \\ 
     & IC~142         &  1 33 55.1  & +30 45 22        & 8.72*   & 7.50*   & 7.42  & 6.78 & KW81 : \\
     & NGC 595        &  1 33 35.50 & +30 41 52        & 8.22*   & 7.00*   & 7.75  & 7.01 & KW81 :\\
     & BCLMP 88 (MA~2)&  1 34 15.5  & +30 37 11        & 8.75*   & 7.50*   & -     & 6.78 & KW81 :\\
     & MA~3           &  1 34 01    & +30 52.1         & 8.45    & 7.10    & 7.63  & 6.75 & KW81 \\ 
     & NGC 604        &  1 34 33.19 & +30 47 05.6      & 8.30    & 6.97    & 7.49  & 6.70 & KW81    \\
     & IC~131         &  1 33 15.0  & +30 45 09        & 8.05    & 6.72    & 7.09  & 6.60 & KW81    \\
     & BCLMP 650 (MA~9a)&1 34 33.64 & +31 00 21.2      & 8.50*   & 7.2*    & 7.33  & 7.10 & KW81 :\\
     & IC~133         &  1 33 15.9  & +30 53 02        & 8.25*   & 6.9*    & 7.55  & 6.95 & KW81 :\\
     & NGC 588        &  1 32 45.2  & +30 38 54.0      & 8.00    & 6.84    & 7.30  & 6.68 & KW81    \\
     & IC 132         &  1 33 15.8  & +30 56 45.0      & 8.06    & 6.72    & 7.39  & 6.70   & KW81    \\
\hline
     & BCLMP 93 (CC93)&  1 33 52.3  & +30 39 18        & 9.02$\pm$0.16    & 7.92   & -  & 7.23$\pm$0.06 & V88:\\
     & IC~142         &  1 33 55.1  & +30 45 22        & 8.70$\pm$0.16    & 7.57   & -  & 7.03$\pm$0.05 & V88 \\
     & NGC 595        &  1 33 35.50 & +30 41 52        & 8.44$\pm$0.09    & 7.33   & -  & 6.86$\pm$0.08 & V88\\
     & BCLMP 88 (MA~2)&  1 34 15.5  & +30 37 11        & 8.44$\pm$0.15    & 7.24   & -  & 6.8$\pm$0.1   & V88\\
     & NGC 604        &  1 34 33.19 & +30 47 05.6      & 8.51$\pm$0.03    & 7.35   & -  & 6.95$\pm$0.03 & V88     \\
     & IC~131         &  1 33 15.0  & +30 45 09        & 8.41$\pm$0.06    & 7.25   & -  & -             & V88    \\
     & NGC 588        &  1 32 45.2  & +30 38 54.0      & 8.30$\pm$0.06    & 6.77   & -  & 6.86$\pm$0.06 & V88    \\
\hline
     &BCLMP 027      &   1 33 45.5  &  30 36 51        & -                &-       &7.68$\pm$0.09&-     & W03\\
     &BCLMP 079      &   1 34 0.2   &  30 40 49        & -                &-        &7.68$\pm$0.04&-     & W03\\
     &BCLMP 087E      &   1 34 2.5   &  30 38 41        & -                &-        &7.96$\pm$0.08&-     & W03\\
     &BCLMP 042      &   1 33 5.5   &  30 39 30        & -                &-        &7.89$\pm$0.10&-     & W03\\
     &BCLMP 301      &   1 33 5.3   &  30 45 22        & -                &-        &7.63$\pm$0.08&-     & W03\\
     &BCLMP 004      &   1 33 9.3   &  30 35 49        & -                &-        &7.85$\pm$0.08&-     & W03\\
     &BCLMP 062      &   1 33 4.5   &  30 44 38        & -                &-        &7.55$\pm$0.08&-     & W03\\
     &BCLMP 049      &   1 33 3.9   &  30 41 28        & -                &-        &7.77$\pm$0.04&-     & W03\\
     &BCLMP 045      &   1 33 8.8   &  30 40 25        & -                &-        &7.66$\pm$0.04&-     & W03\\
     &BCLMP 302      &   1 34 6.9   &  30 47 26        & -                &-        &8.00$\pm$0.04&-     & W03\\
     &BCLMP 214      &   1 33  30   &  30 31 47        & -                &-        &7.84$\pm$0.08&-     & W03\\
     &BCLMP 095      &   1 34 0.9   &  30 36 18        & -                &-        &7.87$\pm$0.08&-     & W03\\
     &BCLMP 088W      &   1 34 5.5   &  30 37 12        & -                &-        &7.72$\pm$0.05&-     & W03\\
     &BCLMP 710      &   1 34 3.6   &  30 33 42        & -                &-        &7.87$\pm$0.07&-     & W03\\
     &BCLMP 702      &   1 34  10   &  30 31 57        & -                &-        &7.84$\pm$0.08&-     & W03\\
     &BCLMP 691      &   1 34 6.4   &  30 51 55        & -                &-        &7.58$\pm$0.04&-     & W03\\
     &BCLMP 680C      &   1 34 2.1   &  30 47  0        & -                &-        &7.70$\pm$0.03&-     & W03\\
     &BCLMP 680B      &   1 34 3.5   &  30 46 50        & -                &-        &7.62$\pm$0.03&-     & W03\\
     &BCLMP 221      &   1 33 9.8   &  30 27 25        & -                &-        &7.65$\pm$0.07&-     & W03\\
     &BCLMP 740W      &   1 34 9.6   &  30 41 52        & -                &-        &7.46$\pm$0.05&-     & W03\\
     &CPDSP 194      &   1 33 1.1   &  30 45 16        & -                &-        &7.53$\pm$0.07&-     & W03\\
     &BCLMP 251      &    1 3336.6  &   30 20 13       & -                &-        &6.00$\pm$0.10&-     & W03\\
     &BCLMP 623      &    1 3316.5  &   30 52 50       & -                &-        &6.90$\pm$0.03&-     & W03\\
     &BCLMP 280      &    1 3245.2  &   30 38 54       & -                &-        &7.30$\pm$0.06&-     & W03\\
     &BCLMP 638E      &    1 3316.3  &   30 56 44       & -                &-        &7.41$\pm$0.08&-     & W03\\
     &BCLMP 638N      &    1 3315.6  &   30 56 49       & -                &-        &7.50$\pm$0.08&-     & W03\\
       \hline
     & BCLMP 090      & 1 34 04.2    & +30 38 09.2       & 8.50$\pm$0.06   & -     & 7.96$\pm$0.09 & -  &C06\\
     & BCLMP 691      & 1 34 16.6    & +30 51 54.0       & 8.26$\pm$0.02   & -     & 7.62$\pm$0.03 & -  &C06\\
     & BCLMP 745      & 1 34 37.6    & +30 34 55.0       & 8.07$\pm$0.10   & -     & 7.2$\pm$0.2   & -  &C06\\ 
     & BCLMP 706      & 1 34 42.2    & +30 31 42.3       & 8.32$\pm$0.12   & -     & 7.6$\pm$0.3   & -  &C06\\ 
     & BCLMP 290      & 1 33 11.4    & +30 45 15.1       & 8.21$\pm$0.06   & -     & 7.4$\pm$0.1   & -  &C06\\ 
     & MA1            & 1 33 03.4    & +30 11 18.7       & 8.24$\pm$0.06   & -     & 7.61$\pm$0.08 & -  &C06\\ 
\hline
     & LGC~HII~2      & 1 32 43.0    & 30 19 31.2        &8.25 $\pm$0.06   &7.17$\pm$0.25 & -      &  6.75$\pm$0.18  & M07 \\	  
     & LGC~HII~3      & 1 32 45.9    & 30 41 35.5        &8.24$\pm$0.05    &6.94$\pm$0.17 & -      &  6.62$\pm$0.11  & M07 \\	  
     & BCLMP289       & 1 32 58.5    & 30 44 28.6        &8.25$\pm$0.13    &6.90$\pm$0.50 & -      &  6.85$\pm$0.3   & M07 \\	  
     & BCLMP218       & 1 33 00.3    & 30 30 47.3        &8.25$\pm$0.05	   &7.19$\pm$0.14 & -      &  6.77$\pm$0.15  & M07 \\	  	 
     & MCM00Em24      & 1 33 10.8    & 30 18 08.5        &8.18$\pm$0.25    &7.10$\pm$0.30 & -      &  6.55$\pm$0.3   & M07 \\	  
     & CPSDP194       & 1 33 11.1    & 30 27 34.2 	 &8.27$\pm$0.06    &7.12$\pm$0.20 & -      &  6.67$\pm$0.17  & M07 \\
     & BCLMP626       & 1 33 16.4    & 30 54 04.8        &8.13$\pm$0.07    &6.86$\pm$0.30 & -      &  6.61$\pm$0.28  & M07 \\
     & BCLMP45        & 1 33 29.0    & 30 40 24.8        &8.49$\pm$0.04    &7.17$\pm$0.15 & -      &  6.92$\pm$0.17  & M07 \\
     & BCLMP637       & 1 33 50.6    & 30 56 33.3        &8.34$\pm$0.05    &7.30$\pm$0.40 & -      &  6.90$\pm$0.40  & M07 \\
     & GDK99~128      & 1 33 59.9    & 30 32 44.3        &8.47$\pm$0.06    &7.31$\pm$0.21 & -      &  6.98$\pm$0.21  & M07 \\
     & BCLMP670       & 1 34 03.3    & 30 53 09.3  	 &8.28$\pm$0.07    &7.14$\pm$0.30 & -	   &  6.75$\pm$0.18  & M07 \\
     & VGHC~2-84      & 1 34 06.7    & 30 48 56.4        &8.35$\pm$0.04    &6.95$\pm$0.18 & -      &  6.56$\pm$0.12  & M07 \\
     & BCLMP717b      & 1 34 37.4    & 30 34 54.3        &8.18$\pm$0.07    &7.04$\pm$0.28 & -      &  6.57$\pm$0.24  & M07 \\
     & LGC~HII~11     & 1 34 42.2    & 30 24 00.5        &8.17$\pm$0.06    &6.68$\pm$0.26 & -      &  6.42$\pm$0.40  & M07 \\
\hline
giant stars&M33 1054  & 1 33 50.61   & 30 38 36.70       & 9.03            &-            &-       &-                  &M97\\  
           &M33 1345  & 1 33 59.676  & 30 23 00.44       & 8.50            &-            &-       &-                  &M97\\  
           &M33 B133  & 1 33 28.848  & 30 47 46.32       & 8.56            &-            &-       &-                  &M97\\  
           &M33 110A  & 1 33 41.023  & 30 22 36.97       & 8.43            &-            &-       &-                  &M97\\  
\hline
           &10900     & 1 33 44.90   &  +30 36 16.70     & 8.30$\pm$0.40   & 8.75$\pm$0.36&-      &-                  &U05\\
           &1110A     & 1 33 41.00   &  +30 22 37.00     & 8.40$\pm$0.40   & 8.50$\pm$0.42&-      &-                  &U05\\
           &1B38      & 1 33 00.83   &  +30 35 05.10     & 8.15$\pm$0.29   & 8.50$\pm$0.20&-      &-                  &U05\\
           &1B133     & 1 33 29.00   &  +30 47 44.00     & 8.10$\pm$0.16   & 8.30$\pm$0.17&-      &-                  &U05\\
           &11054     & 1 33 50.83   &  +30 38 34.50     & 8.10$\pm$0.15   & 8.90$\pm$0.18&-      &-                  &U05\\
           &11137     & 1 33 53.23   &  +30 35 26.10     & 8.50$\pm$0.30   & 8.70$\pm$0.25&-      &-                  &U05\\
           &1OB11241  & 1 33 42.02   &  +30 21 42.30     & 8.00$\pm$0.20   & 8.60$\pm$0.15&-      &-                  &U05\\
           &1UIT103   & 1 33 27.29   &  +31 00 56.70     & 7.90$\pm$0.10   & 8.20$\pm$0.13&-      &-                  &U05\\
           &1UIT122   & 1 33 33.70   &  +30 47 20.20     & 8.10$\pm$0.12   & 8.50$\pm$0.06&-      &-                  &U05\\
           &1UIT136   & 1 33 35.73   &  +31 00 47.00     & 7.70$\pm$0.19   & 8.40$\pm$0.18&-      &-                  &U05\\
           &1OB10-10  & 1 33 44.22   &  +30 31 48.20     & 8.55$\pm$0.21   & 8.65$\pm$0.22&-      &-                  &U05\\
\hline
 Cepheids  &A121,029  & 1 34 59.72   &	30 52 25.2       &8.22 	           &-            &-       &-                  &B06\\ 
           &B160,520  & 1 32 56.82   &	30 41 33.8	 &8.30 	           &-            &-       &-                  &B06\\ 
           &C133,449  & 1 34 33.43   &	30 51 15.6	 &8.56	           &-            &-       &-                  &B06\\ 
           &D234,922  & 1 33 54.63   &	30 35 19.8	 &8.57 	           &-            &-       &-                  &B06\\ 
           &E237,367  & 1 34 03.97   &	30 38 08.4	 &8.58             &-            &-       &-                  &B06\\ 
\hline  
PNe        &PN8       & 1 32 28.66   &  30 25 53.2       &  7.94           &7.07         &-        & 6.38              &M04\\
           &PN18      & 1 33 06.11   &  30 31 04.5       &  8.20           &7.48         &-        & 6.83              &M04\\
           &PN28      & 1 33 19.25   &  30 29 40.4       &  8.69           &6.98         &-        & 6.47              &M04\\
           &PN60      & 1 33 46.42   &  30 26 55.3       &  8.69           &7.67         &-        & 6.57              &M04\\
           &PN65      & 1 33 49.37   &  30 32 06.3       &  8.67           &7.61         &-        & 6.84              &M04\\
           &PN75      & 1 34 01.14   &  30 50 27.1       &  8.49           &8.03         &-        & 6.81              &M04\\
           &PN91      & 1 34 13.97   &  30 22 36.4       &  8.48           &8.07         &-        & 6.60              &M04\\
           &PN93      & 1 34 15.88   &  30 24 54.6       &  8.47           &7.75         &-        & 6.66              &M04\\
           &PN96      & 1 34 15.44   &   30 32 20.2      &  8.76           &7.43         &-        & 6.44              &M04\\
           &PN101     & 1 34 24.25   &   30 27 54.3      &  8.72           &7.33         &-        & 6.18              &M04\\
           &PN125     & 1 34 32.85   &  30 41 10.5       &  8.31           &7.53         &-        & 6.66              &M04\\
\end{longtable}
}

\end{document}